\documentclass[pra,aps,twocolumn,longbibliography, nofootinbib, showpacs]{revtex4-2}
\usepackage[colorlinks=true, citecolor=blue, urlcolor=blue, linkcolor = blue ]{hyperref}
\usepackage{epsfig,newlfont,amssymb,amsfonts,amsmath,bm,subfigure,palatino,mathtools,amsthm,braket,times,soul,enumitem,color}
\usepackage[normalem]{ulem}
\newcommand{\stkout}[1]{\ifmmode\text{\sout{\ensuremath{#1}}}\else\sout{#1}\fi}
\usepackage[T1]{fontenc}
\usepackage[normalem]{ulem}
\usepackage{graphicx}
\usepackage{placeins}
\usepackage{hyperref}
\usepackage{multirow}
\usepackage{hhline,booktabs}
\usepackage{xcolor}
\usepackage{bbm}

\DeclareMathAlphabet{\mathpzc}{OT1}{pzc}{m}{it}

\begin{document}

\title{Realizing quantum energy storage and sensing with prethermal time crystals stabilized via power-law-graded Ising interactions}


\author{Ayan Sahoo and Debraj Rakshit}
\affiliation{Harish-Chandra Research Institute, Chhatnag Road, Jhunsi, Prayagraj 211 019, India \\
Homi Bhabha National Institute, Training School Complex, Anushakti Nagar, Mumbai
400 094, India}

\begin{abstract}
 We study {\color{black}{prethermal discrete time-crystalline (PDTC)}} phases in one-dimensional spin-1/2 chains with power-law–modulated nearest-neighbor interactions under periodic Floquet driving.  This system supports robust period-doubled dynamics across a wide range of grading exponents, stabilized by the interplay between coherent driving and spatially varying coupling. This work proposes the PDTC phases as an excellent platform for engineering robust quantum batteries. Within the PDTC phase, the energy stored in the system, interpreted as a quantum battery, increases superlinearly with system size, although no super-liner scaling advantage persists in the normalized stored energy. {\color{black}{The main advantage of the PDTC phase in quantum batteries arises from emergent stable subharmonic responses by suppression of heating.}} These responses remain stable even in the presence of imperfections in initial state preparation or deviations from ideal driving. As a result, the PDTC-based approach is inherently offers robust platform for realizing quantum battery. In contrast, standard time-dependent protocols require careful parameter fine-tuning or precise synchronization of measurements with the optimal time for extracting maximal power during transient dynamics. Beyond energy storage, we demonstrate that the PDTC phase supports enhanced quantum sensing in a wide range of the gradient exponent. {\color{black}{The quantum Fisher information associated with estimating timing deviations in the drive scales superextensively with system size, indicating quantum enhanced sensitivity.}} Particularly, the degree of quantum advantage can be tuned by varying the interaction exponent, though PDTC behavior remains robust throughout. Our results position power-law–graded interacting Floquet systems as robust platforms for storing quantum energy and achieving metrological enhancement.

\end{abstract} 

\maketitle

{\color{black}

\section{Introduction}

Spontaneous symmetry breaking (SSB) is one of the most fundamental ideas in modern physics, which describes various phases of matter and can identify the phase transition point of different states of matter. For example, the crystal forms due to the breaking of continuous spatial translational symmetry. Expanding this idea into the context of time, Frank Wilczek first introduced the idea of discrete-time translational symmetry breaking, which opens a way for the theoretical exploration of time crystals. The time-crystal system show motion or oscillation with higher periods with integer multiples of the external driving period \cite{Wilczek2012}. The quantum system subjected to a periodic drive, known as a floquet system, can exhibit discrete-time translational symmetry breaking, meaning that the system behaves in a way that its observables oscillate with a period which is greater than and an integer multiple of the driving period \cite{Li2012, Wilczek2013, PhysRevX.12.031037, PhysRevLett.120.215301}. Discrete time-crystalline (DTC) phase has been experimentally probed in programmable quantum materials \cite{Choi_Nature_17, hess_Nature_17, Zhan_Nature_22, Frey2022, Autti2020}.

In a periodically driven floquet system, the energy is not conserved as a result of the continuous injection of energy due to the external drive. An isolated non-integrable quantum system is then expected to absorb energy indefinitely, and eventually heat up to the featureless infinite temperature state at long times. However, integrability can lead to long-time steady states \cite{Serbyn2013, Huse2014, Fleckenstein2021, Ho2023, Mishra2019}.
Another route for avoiding the heat up to the featureless infinite temperature state is through disorder-induced many-body localization (MBL), which prevents thermalization to the infinite-temperature state by suppressing energy absorption \cite{Khemani2016,Else2016, von2016, Ponte2015, von2016b, PhysRevB.111.024315}. This stabilization preserves the unique properties of the time crystals over time. 
The applications of time crystals are numerous, including the design of quantum engines \cite{PhysRevLett.125.240602}, improving quantum computation through error-resistant topological properties \cite{Bomantara2018}, and simulating intricate quantum systems \cite{science_adv_20}. They provide platform for studying the effects of AC fields, system-environment interactions \cite{Iemini2024, Cabot2024}.

Of particular relevance to the present study is Ref.~\cite{Liu2023}, which introduces a disorder-free route to stabilizing nontrivial DTC phase via Stark many-body localization in a periodically driven setting. Stark localization, in its conventional form, considers a lattice subjected to a linear onsite potential that causes suppression of transport. In this work, we 
generalize this framework to one-dimensional spin-1/2 chains with power-law–graded interactions subjected to periodic Floquet driving. We show that such systems exhibit robust period-doubled dynamics characterizing the PDTC order across a wide range of interaction exponents. The robustness is attributed to the cooperative effects of the external driving and spatially varying interaction profile. {{\color{black}{Although a detailed characterization of the underlying mechanism for this model is beyond the scope of the present work, it presents a promising avenue for future studies \cite{debnath2025localization}.}} 

Beyond their fundamental interest, we demonstrate that the PDTC phases stabilized via generalized power-law-graded interaction have potential applications in emerging quantum technologies. In particular, the long-lived subharmonic response and robustness against perturbations can be harnessed for quantum batteries, where periodic driving protocols can be utilized for coherent energy storage. 
We propose that PDTC phases provide a fundamentally new route to harness nonequilibrium quantum matter for applications such as energy storage and information processing. Unlike conventional quench dynamics, which typically lead to transient coherent oscillations that eventually decay, or standard Floquet engineering, where the long-time dynamics are sensitive to heating and parameter fine-tuning, the {\color{black}{PDTC phase offers robustness through emergent subharmonic responses that persist for sufficiently long times compared to operational quantum battery timescale.}} In particular, the ability of a PDTC to lock oscillations at a frequency commensurate with the drive, regardless of microscopic details, opens a pathway toward stable and scalable protocols where energy storage and retrieval can be maintained over long times without fine control of external driving fields.

Similarly, the inherent phase stability of PDTC dynamics can be exploited for quantum sensing, where the period-doubled oscillations can serve as a means to detect weak perturbations through shifts in the oscillation period or amplitude \cite{Iemini2024, Yousefjani2025}. Previous proposals have explored sensing based on the PDTC phase, demonstrating resilience to noise and long coherence times. Particularly, our work draws inspiration from the sensing framework presented in \cite{Yousefjani2025}, where discrete time-crystal dynamics were exploited for enhanced sensitivity. Our approach complements and extends those ideas by considering a platform with the power-law–graded interaction.

In principle, this intrinsic stability against imperfections and perturbations makes the PDTC phase an attractive platform for realizing reliable quantum devices. Moreover, a versatile platform for both energy storage and precision metrology can be realized by tuning the interaction ranges in the system considered in this work. A more detailed discussion of the relevant literature and reviews on quantum batteries and sensing applications will be presented in the respective sections of this work.

This paper is presented in the following sections. Section I presents the introduction. In Sec .~II, we introduce the model with the idea of how the external driving field is implemented through floquet dynamics. The behaviour of the energy spectrum of the system is also described here. Section III examines the time crystaline behaviour of our system across the different parameter regimes. In this section, we analytically show that the quasi energy of the system forms a $\pi$-pair in the PDTC phase.  Section IV explores the application of time crystals as a quantum battery. Then in the Sec.~V, we describe the advantages of quantum sensing by utilizing the PDTC phase and how the performance of the sensor depends on the grading exponent of the local interaction strength. Finally, the conclusion is provided in Sec.~VI.

\section{Model}
\label{model}

We consider a spin-$1/2$ chain of length $L$ with $z$–$z$ interactions, subject to open boundary conditions. The interaction strength increases as $j^{\alpha}$, where $j$ denotes the site index along the chain and $\alpha$ is a tunable parameter. The Hamiltonian of this system is described as 
\begin{align}
    \label{eq:battery}
     H_B&= \sum_{j = 1}^{L-1} j^{\alpha} \sigma_z^j \sigma_z^{j+1},
\end{align}
where $\sigma_{z}^{j}$ is $z$-component of Pauli operator of $j$'th position, and the exponent $\alpha$ controls the spatially varying interaction profile. The system is driven by a time-dependent field $V(t)$ which periodically acts on the system. The modified Hamiltonian now becomes 
 \begin{align}
    \label{eq:charger}
     H_c&= H_B + V(t), 
    \end{align}
where the time-dependent driving term is
\begin{align}
    \label{eq:V(t)}
    V(t)&= \sum_n \delta(t-nT)(\Phi H_{\text{kick}}).
\end{align}
Here, $\delta(t - nT)$ denotes the Dirac delta function, ensuring that the kicks act instantaneously at times $t = nT$ with $T$ as time interval, and $\Phi$ is a dimensionless parameter controlling the kick strength. The $\Phi = \frac{\pi}{2}(1-e)$ is the dimensionless parameter controlling the strength of the kick Hamiltonian $H_{\text{kick}}$,  where $e$ is the small perturbation over to the ideal delta kick. The kick Hamiltonian is given by
 \begin{align}
    \label{eq:kick}
  H_{kick} = \sum_j  \sigma_x^{j}.
 \end{align}
 where $\sigma_{x}^{j}$ is the $x$-component of the Pauli operator acting on $j$'th site. 
 In this work we consider cases where either all of the spins experience the delta kick or the alternative ones. The stroboscopic dynamics of the system can be described by the Floquet operator $U_F$, which represents the time-evolution operator over one driving period. 
{\color{black}{In this work, we consider two different kinds of Floquet protocols depending on the rotation induced by the delta kick: the first kind --
\begin{align}
     \label{eq:floquet_operator}
 {U}_F = e^{-i\Phi\sum_{j=1}^{L}  \sigma_x^{j}} e^{-i T H_B}, 
\end{align}
where all spins undergo rotation, and a second kind -- 
\begin{align}
\label{eq:floquet_operator_2}
{\tilde U}_F = e^{-i\Phi\sum_{j=1}^{L/2}  \sigma_x^{2j}} e^{-i T H_B},
\end{align}
where only the alternate spins are rotated. {\color{black}{We have chosen $T=\pi/(2L)$. In a realistic experimental setup with a finite pulse width $T'>0$, the total Floquet period remains finite as $L\to\infty$. Specifically, as the free-evolution time $T=\pi/(2L)$ vanishes in this limit, the total driving period approaches the finite pulse duration $T'$.}} In our study, we employ different forms of the Floquet operator depending on the physical advantage required. For instance, to achieve maximum energy storage, we incorporate the alternating-rotation Floquet operator (see Sec.~\ref{sec:battery}).}} 

{\color{black}{ Tunable interaction is experimentally accessible in current experimental platforms, such as trapped ions and Rydberg atom arrays. In a linear chain of trapped ions, effective spins can be encoded in the hyperfine levels, while Raman laser beams generate phonon-mediated long-range spin–spin interactions with a tunable power-law form. Therefore, trapped-ion platforms provide a realistic and controllable route for implementing the power-law-graded interacting Floquet system.  In an ion trap, a selective "kick" on chosen ions can be implemented by a tightly focused Raman beam, so that only the addressed ions experience the short Raman pulse that flips their spins \cite{PhysRevLett.92.207901, Islam2013, RevModPhys.93.025001}.}}

We first note that $H_B$ is diagonal in the computational basis, defined as the set of tensor-product states  $\{\,|x\rangle\,\} = \{\,|\sigma_1, \sigma_2, \dots, \sigma_L\rangle \,\}$, where $\sigma_j \in \{\uparrow, \downarrow\}$ denotes the spin state at site $j$. The single-spin states $|\uparrow\rangle$ and $|\downarrow\rangle$ are eigenstates of the Pauli operator $\sigma_z$, such that $\sigma_z |\uparrow\rangle = +\,|\uparrow\rangle$ and $\sigma_z |\downarrow\rangle = -\,|\downarrow\rangle$. $H_B$ can be re-expressed in the computational basis as $H_B = \sum_{x=1}^{2^L} E_x |x\rangle \langle x|$,
where $E_x$ is the energy eigenvalue associated with the eigenstate $|x\rangle$. 

The Hamiltonian $H_B$ exhibits two discrete $\mathbb{Z}_2$ symmetries that constrain its spectral properties. (i) Global spin-flip symmetry is defined via the unitary operator $F = \prod_{j=1}^L \sigma_x^j$, which flips all spins in the $z$-basis according to $\sigma_x |\uparrow\rangle = |\downarrow\rangle$ and $\sigma_x |\downarrow\rangle = |\uparrow\rangle$. Under this transformation, each $\sigma_z^j$ changes sign, $F \, \sigma_z^j \, F^{-1} = - \sigma_z^j$, so that each product $\sigma_z^j \sigma_z^{j+1}$ remains invariant: $(-\sigma_z^j)(-\sigma_z^{j+1}) = \sigma_z^j \sigma_z^{j+1}.$ Thus $[F, H_B] = 0,$ and $F$ is a symmetry of $H_B$. This invariance implies that every eigenstate $|x\rangle$ has a partner $F|x\rangle$ with the same energy, leading to at least a twofold degeneracy unless $|x\rangle$ is itself invariant under $F$.  (ii) The second one is the sublattice spin-flip symmetry, which  for a bipartite chain with open boundaries is understood via $P = \prod_{j=1}^{L/2} \sigma_x^{2j}$, which flips the spins only on one sublattice (e.g., all even sites). In this case, one spin in each nearest-neighbor pair is flipped while the other remains unchanged, giving $P \, \sigma_z^j \sigma_z^{j+1} \, P^{-1} = - \sigma_z^j \sigma_z^{j+1}$. Consequently, $P H_B P^{-1} = - H_B,$ which means $P$ anticommutes with $H_B$, i.e., $\{ P, H_B \} = 0.$ This anticommutation enforces a spectral symmetry: if $|E\rangle$ is an eigenstate of $H_B$ with eigenvalue $E$, then $P|E\rangle$ is an eigenstate with eigenvalue $-E$. The spectrum is therefore symmetric about zero energy. Both symmetries persist for arbitrary values of the exponent $\alpha$ in the coupling profile. However, the introduction of either a longitudinal field term $\sum_j h_z \sigma_z^j$ or a transverse field term $\sum_j h_x \sigma_x^j$ breaks one or both of these symmetries, thereby lifting the associated degeneracies or spectral symmetry.

\begin{figure*}[t!]
    \centering
    \includegraphics[scale = 0.42]{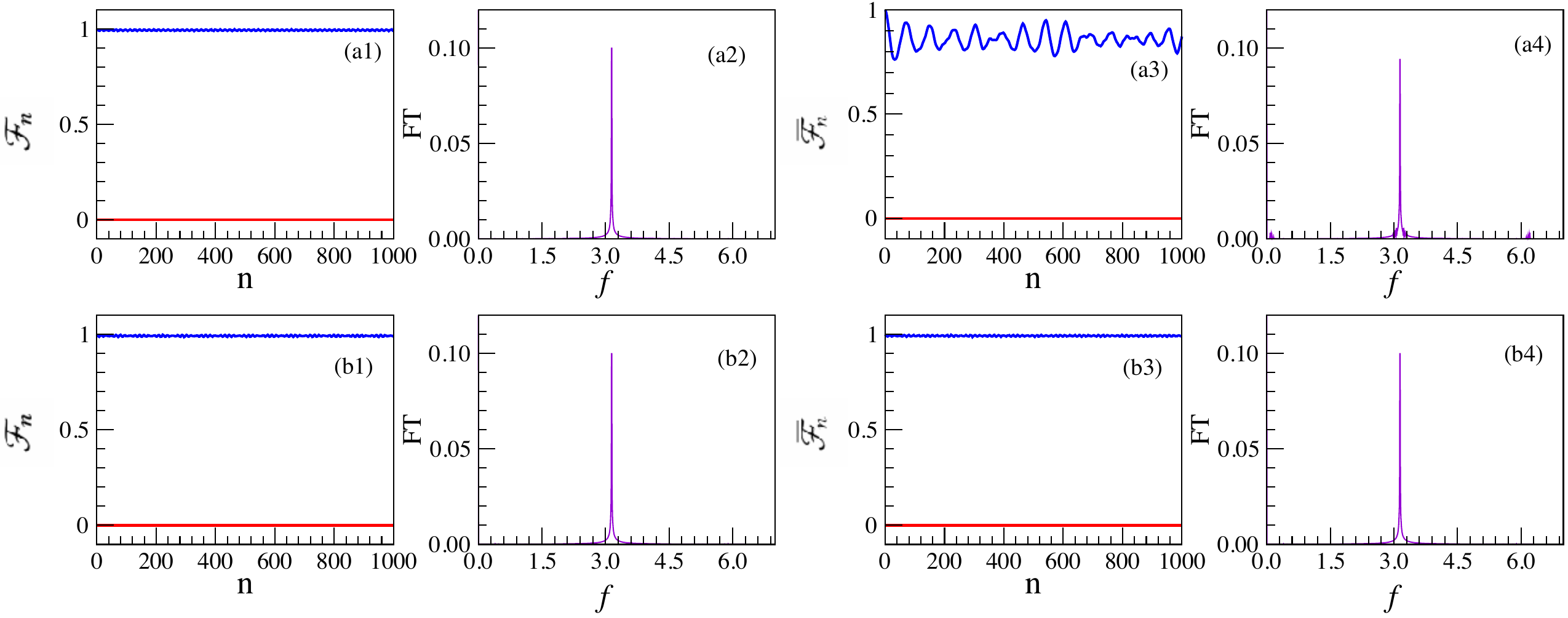}
    \caption{ {\bf For value \bm{$\alpha = 0.5$} :} {\bf (a1)} The plot presents the fidelity $\mathcal{F}_n$ (as defined in Eq.~\ref{fidelity}) along y-axis  with the function of $n$ which is along x-axis for the parameter value $\alpha = 0.5$ for the specific initial state $|x_{\text{min}}\rangle$. The blue and red color line represents the fidelity for even and odd $n$, respectively. For even $n$, the $\mathcal{F}_n$ is $1$, while for odd $n$, the $\mathcal{F}_n$ is close to zero, indicating the system has period doubling in dynamics. The corresponding Fourier transform (FT) is shown in  {\bf (a2)}, where a prominent peak is observed near $\pi$, which further confirms the period doubling of $\mathcal{F}_n$. {\bf (a3)} This plot is the average value of all product state fidelity $\bar{\mathcal{F}}_n$ (defined in Eq.~\ref{average fidelity}) with $n$, and the corresponding FT is given in the Figure {\bf (a4)} which also confirms the period doubling by peak at $\pi$. This confirms the robustness of period doubling with initial states.  A similar analysis have done for parameter value {\bf\bm{$\alpha = 1.5$} :} {\bf (b1)} This figure depicts the $\mathcal{F}_n$ versus $n$ for the same initial state. The blue color line and red color line again display that $\mathcal{F}_n$ is $1$ and zero, respectively. and the plot {\bf (b2) } is FT of {\bf (b1)} also show peak at $\pi$. The Figure {\bf (a4)} is for the $\bar{\mathcal{F}}_n$  versus  $n$ for $\alpha = 1.5$. The corresponding FT, which also peaks at $\pi$, shows in {\bf (b4)}. This confirms that even at higher values of $\alpha$ the system shows period doubling in the PDTC phase. All of the above figures are presented for system size $L=8$. In the Fourier transform plot, the x-axis represents the frequency $f$.  }
    \label{fig:timecrystal}
\end{figure*}

\begin{figure*}[t!]
    \centering
    \includegraphics[scale = 0.4]{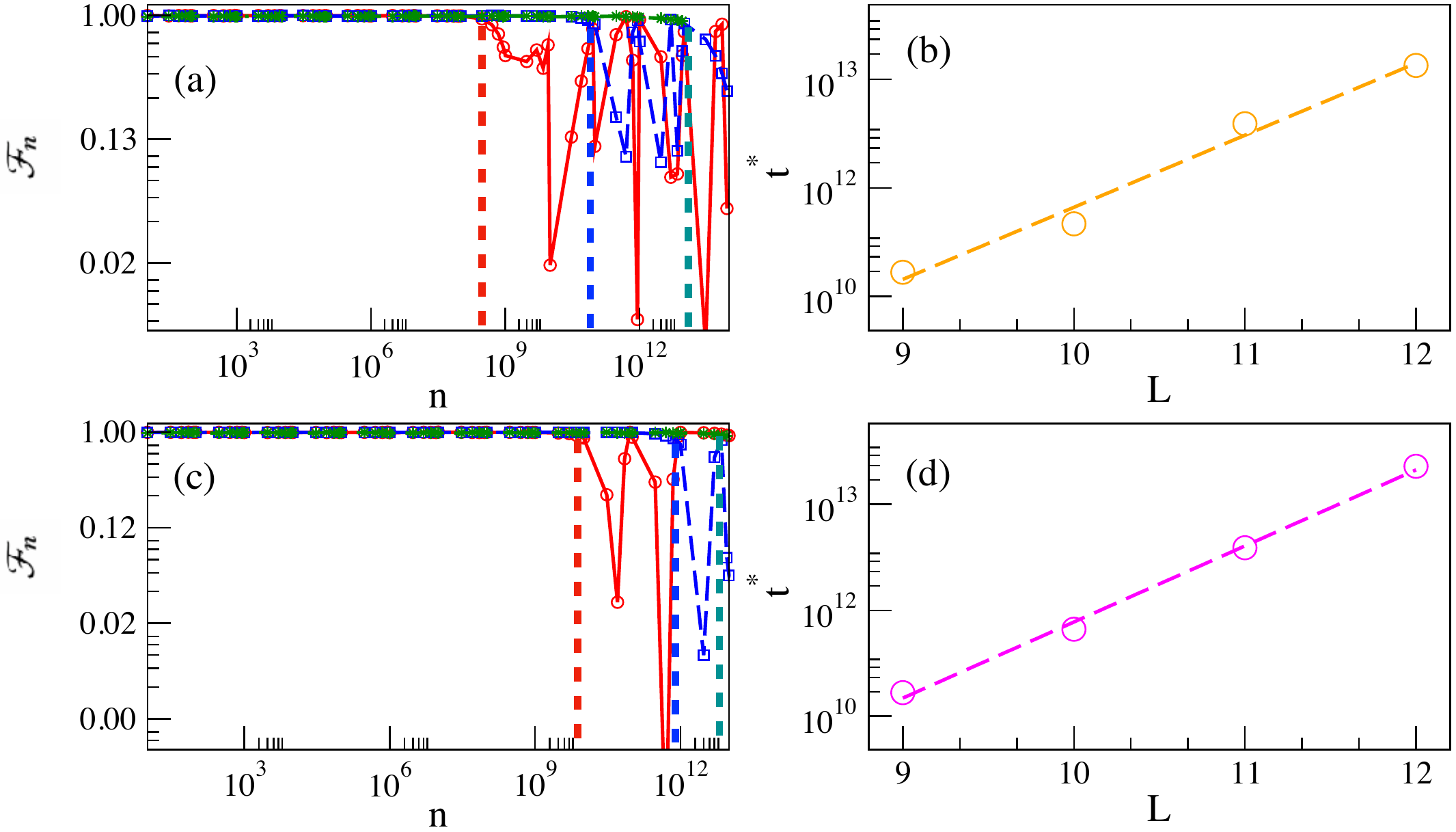}
    \caption{  {\color{black}{ {\bf (a)} This panel shows the fidelity at long stroboscopic times for parameters $e = 0.01$ and $\alpha = 0.5$, for system sizes $L = 6$ (red), $L = 8$ (blue), and $L = 12$ (green) with the initial state $|\uparrow\downarrow\uparrow\downarrow\dots\rangle$. {\bf (b)} presents the scaling of the DTC lifetime $t^{*}$ with system size $L$ and $\alpha = 0.5$. The lifetime increases exponentially with $L$. The orange circles represent the extracted values of $t^{*}$ for different system sizes, and the orange dotted line shows the fitting function $t^{*} \sim e^{1.99(2)L}$.  {\bf (c)}  A parallel plot of the fidelity $\mathcal{F}_n$ versus the stroboscopic time $n$ is shown for $\alpha = 1.5$, using the same parameters as in panel (a). {\bf (d)} shows the scaling of the DTC lifetime $t^{*}$ with system size for $\alpha = 1.5$. The magenta circles represent the extracted values of $t^{*}$, and the magenta dotted line corresponds to the fitting function $t^{*} \sim e^{2.15(9)L}$. The vertical dotted lines in panels (a) and (c) mark the maximum value of $n$ for which the system is in the PDTC phase. The color scheme represents the corresponding system size associated with $t^{*}$.
 } }}
    \label{fig:life_time_plot}
\end{figure*}

\section{Prethermal Time crystal}
\label{timecrystal}



{\color{black}{The PDTC phase is marked by following key features: First, the system shows a subharmonic response that is characterized by oscillations of a period $gT$, where $g$ is an integer greater than one and $T$ is the time period of the external driving field. This subharmonic response corresponds to the spontaneous breaking of the time-translational symmetry imposed by the driving field. Second, these subharmonic responses are remarkably robust against small imperfections in the driving field or system parameters, indicating that the observed dynamics is an intrinsic property of the system. Third, the oscillations persist for a long but finite prethermal time scale before eventually decaying due to Floquet heating. Taken together, the robust subharmonic response and its long-lived persistence over the prethermal time window characterize the system as a PDTC.
} }

{\color{black}Our system has a subharmonic response of period doubling for both the rotational protocols introduced in Eqs.~(\ref{eq:floquet_operator}-\ref{eq:floquet_operator_2}). In this section, we explicitly provide a detailed discussion on the emergence of the crystaline phase due to the Floquet protocol in Eq.~(\ref{eq:floquet_operator}), which is utilized for designing a time-crystal-based quantum sensor (Sec.~\ref{sec:sensing}). The emergence of quasienergy pairs separated by $\pi$  is a characteristic of period doubling. Let us consider the Floquet operator $U_F^0$ at $e=0$, whose spectral decomposition is given by 
\begin{equation}
U_F^0 = \sum_k e^{-i\phi_k} |\phi_k\rangle \langle \phi_k|,
\end{equation}
where $\phi_k$ refers to the quasienergy and $|\phi_k\rangle$ is the corresponding Floquet eigenstate (also referred to as a quasi-eigenstate). In the context of PDTC dynamics, the Floquet eigenstates can take the form
\begin{equation}
|\phi^{\pm}\rangle = \frac{1}{\sqrt{2}} \left( e^{-i E_{-x}T/2} |x\rangle \pm e^{-i E_x T/2} |-x\rangle \right),
\end{equation}
where $|x\rangle$ is a product state defined previously in Sec.~\ref{model}, and $|-x\rangle = \prod_{j=1}^{L} \sigma_x^j |x\rangle$ is obtained by applying a global spin-flip operation to $|x\rangle$. After acting floquet operator on the states $|\phi^{\pm}\rangle$, the final state will be 
\begin{equation}
\label{eq:quasienergy_pair}
U_F^0 |\phi^{\pm}\rangle = \pm e^{-i(E_x + E_{-x})T/2} |\phi^{\pm}\rangle.
\end{equation}
Now let's assume $E = (E_x + E_{-x})T/2$ as the quasienergy associated with $|\phi^+\rangle$. Then the final states after applying $U_F^{0}$ become 
\begin{align}
 U_F^0 |\phi^+\rangle &= e^{-iE} |\phi^+\rangle, \nonumber \\       
 U_F^0 |\phi^-\rangle &= -e^{-iE} |\phi^-\rangle = e^{-i(E + \pi)} |\phi^-\rangle. 
\end{align}
The above relations clearly show that the quasienergies associated with the states $|\phi^+\rangle$ and $|\phi^-\rangle$ differ by exactly $\pi$ for $e=0$. This $\pi$ pairing of the quasi-energies is a hallmark of DTC phases with period doubling, i.e., the system returns to its initial state after twice the period of the driving field. {\color{black}{However, for finite $e$, we also analyze the stability of the $\pi$-pairing (Appendix~\ref{appendix_timecrystal}). For small values of $e$, the $\pi$-pairing remains robust across the entire quasienergy spectrum, whereas for larger $e$, the $\pi$-pairing breaks down in a part of the spectrum. }}   }

{\color{black} {In the absence of perturbation, i.e., $e = 0$, our system exhibits period doubling through spontaneous breaking of time-translation symmetry for any initial product states $|x\rangle$. In order to illustrate this, we consider a specific initial state $|x\rangle = |\uparrow \downarrow \uparrow \downarrow \cdots\rangle$ and compute the evolved state after $2n$ stroboscopic periods, which is given by
\begin{equation}
(U_F^0)^{2n} |x\rangle = e^{-i 2n (E_{x} + E_{-x}) T} |x\rangle,
\end{equation}
where $E_{x}$ and $E_{-x}$ denote the eigenenergies associated with the states $|x\rangle$ and its orthogonal partner $|-x\rangle = \prod_{j=1}^{L} \sigma_x^j |x\rangle$, respectively. So, the final evolved state returns to its initial state after $2n$ period up to a global phase. The fidelity after $2n$ periods, 
\begin{equation}
\mathcal{F}^{2n}(e=0, T) = |\langle x | (U_F^0)^{2n} | x \rangle|,
\end{equation}
which evaluates to 1, as expected. This result confirms the presence of period doubling and the breaking of discrete time-translation symmetry in the system under consideration, when $e=0$. On the other hand, after acting odd number of driving periods, as example after a single drive period, the state is $U_F^{0} |x\rangle = e^{-i  E_{x} T} |-x\rangle$, indicating that evolution from $|x\rangle$ to an orthogonal state $|-x\rangle$. As a result, the fidelity vanishes after odd numbers of drives, i.e.,
\begin{equation}
\mathcal{F}^{2n+1}(e=0, T) = |\langle x | (U_F^0)^{2n+1} | x \rangle| = 0.
\end{equation}

}}
{\color{black}This alternating pattern, unit fidelity for even $n$ and vanishing fidelity for odd $n$, directly follows from the $\pi$-pairing of quasienergies enforced by the sublattice spin-flip symmetry, and this cycle persists indefinitely for $e=0$, and throughout the prethermal window, $t \le t^* \sim e^{cL}$, without decay for $e \ne 0$, signaling a PDTC}. However, confirming its stability requires testing the robustness of this behaviour against perturbations that break the underlying symmetry.

In the presence of a driving imperfection, quantified by the perturbative parameter $ e\neq 0$, we analyze the system numerically for large $L$. In this case, the floquet operator is modified to 
\begin{equation}
U_F = e^{ie\frac{\pi}{2}\sum_{j=1}^{L}  \sigma_x^{j}} U_F^0,
\end{equation}
where $U_F^0$ is the Floquet operator in the unperturbed case. The exponential prefactor represents a uniform rotation of each spin about the $x$-axis by an angle $e\pi/2$, thus introducing a non-trivial perturbation to the ideal dynamics.

For the case of a small system, $L=4$ and $n = 2$, where a fully analytic treatment is possible, we provide an explicit expression for the fidelity for an arbitrary initial product state $|x\rangle$ (see Appendix~\ref{appendix_timecrystal} for the derivation).
\begin{align}
\langle x | U_F^{2} | x \rangle &= (-i)^{2L}e^{-2E_x T} \sum_{n_f} \left[\cos(\phi)^{2(L-n_f)} \left(i\sin(\phi)\right)^{2 n_f} \right] \nonumber\\
   & \quad \times \sum_{\{n_f\}}e^{-iE_{-x}\{n_f\} T},
\end{align}
where $\{n_f\}$ denotes the set of all possible spin-flip configurations corresponding to $n_f$ number of rotations applied to the state $|-x\rangle$, and $E_{-x}\{n_f\}$ represents the energy of the resulting configuration after these $n_f$ spin flips.

We numerically demonstrate that the system shows robust time-crystalline behavior at the particular driving period $T = \pi / 2L$. To investigate this phenomenon in greater detail, we present numerical results for two representative values of $\alpha$, namely $\alpha = 0.5$ and $\alpha = 1.5$. To quantitatively characterize the time-crystalline behavior, we evaluate two distinct fidelity measures. The first is the fidelity for a chosen initial state,
\begin{equation}
 \label{fidelity}
    \mathcal{F}_n = |\langle \psi_i | U_F^{n} | \psi_i \rangle|^2,
\end{equation}
where $|\psi_i\rangle$ represents the initial state of the system. Second, we compute the state-averaged fidelity, defined as
\begin{equation}
\label{average fidelity}
\overline{\mathcal{F}}_n = \frac{1}{2^L} \sum_{\{x\}} |\langle x | U_F^{n} | x \rangle|,
\end{equation}
which gives an average measure across all possible product states,  $\{|x\rangle\}$, as the initial state. In the Fig. ~\ref{fig:timecrystal}(a1), we present the fidelity $\mathcal{F}_n$ as a function of $n$ for the initial state $|x_{\text{min}}\rangle=|\uparrow\downarrow\uparrow\downarrow\ldots\rangle$. The blue color represents the  $\mathcal{F}_n$ for even values of $n$ and the red color corresponds to the odd values of $n$ for $\alpha=0.5$, $e=0.01$, and $L=8$. This result displays period-doubling behavior of the system that returns to its initial state every even $n$, with vanishing fidelity at odd $n$. Furthermore, this period doubling is confirmed by the Fourier transform (FT) shown in Fig.~\ref{fig:timecrystal}(a2), which has a peak at frequency $\pi$.

The confirmation of the time crystalline phase also demands that the observed behaviour is independent of any specific choice of the initial state. In order to take into account all other initial states, we present the state-averaged fidelity $\overline{\mathcal{F}}_n$ for $\alpha = 0.5$, $e=0.01$, and $L=8$ in Fig.~\ref{fig:timecrystal}(a3). We find that the value of $\overline{\mathcal{F}}_n$ for even $n$ (blue curve) remains close to unity, while it nearly vanishes for odd $n$ (red curve), indicating robust period doubling. The corresponding FT in Fig.~\ref{fig:timecrystal}(a4) reveals a pronounced peak at frequency $\pi$, confirming that the period doubling behavior is independent of the choice of the initial state.

A parallel analysis for $\alpha = 1.5$ is presented, in the Fig.~\ref{fig:timecrystal}(b1), where $\mathcal{F}_n$ is plotted against $n$ for $e=0.01$, $L=8$, and $\alpha = 1.5$. Here also the fidelity $\mathcal{F}_n$ for the same initial state demonstrates that the system reverts to its initial state at even $n$ (blue color), with $\mathcal{F}_n = 1$ and $\mathcal{F}_n = 0$ at odd $n$ (red color). The corresponding FT has a sharp peak at $ \ pi$ again (see Fig.~\ref{fig:timecrystal}(b2)). The state-averaged fidelity for $\alpha = 1.5$ is displayed in Fig.~\ref{fig:timecrystal}(b3). It follows the same trend, remaining close to unity at even $n$ and nearly zero at odd $n$. Its FT in Fig.~\ref{fig:timecrystal}(b4) confirms the period doubling for $\alpha = 1.5$ as well. These results demonstrate that the time-crystalline behavior persists across different values of $\alpha$ even in the presence of a small perturbation $e$. \\

{\color{black}{  The long-time behavior of $\mathcal{F}_n$ is presented in Fig.~\ref{fig:life_time_plot}. In Fig.~\ref{fig:life_time_plot}(a), we plot $\mathcal{F}_n$ as a function of $n$ over long evolution times for different system sizes $L = 6$ (red circles), $L = 8$ (blue squares), and $L = 12$ (green diamonds), while keeping the parameters fixed at $\alpha = 0.5$ and $e = 0.01$. From the figure, it is clearly observed that $\mathcal{F}_n$ remains nearly constant up to a certain value of $n$, beyond which it starts exhibiting pronounced oscillations. The regime in which $\mathcal{F}_n$ maintains a constant value corresponds to the PDTC phase, indicating a robust subharmonic response persisting over a long time. In contrast, the emergence of oscillations signals the breakdown of the DTC order and the transition to a non-DTC phase. We define the lifetime, denoted by $t^{*}$, as the maximum value of $n$ up to which $\mathcal{F}_n$ remains constant before the onset of oscillatory behavior. The dependence of the PDTC lifetime on system size is shown in Fig.~\ref{fig:life_time_plot}(b). In this panel, the orange circles represent the numerically extracted values of $t^{*}$ for system sizes $L = 9, 10, 11, 12$. The orange dotted line corresponds to an exponential fit of the lifetime as a function of system size, given by $t^{*} \sim e^{1.99(2)L}$. This fitting clearly demonstrates that the PDTC lifetime increases exponentially with increasing system size, indicating that the stability of the PDTC phase is significantly enhanced in larger systems.

A similar analysis is carried out for $\alpha = 1.5$, as presented in Fig.~\ref{fig:life_time_plot}(c). In this case, $\mathcal{F}_n$ exhibits behavior qualitatively similar to that observed for $\alpha = 0.5$, with an initial regime where $\mathcal{F}_n$ remains approximately constant, followed by a transition to an oscillatory regime at longer times. Accordingly, the PDTC lifetime $t^{*}$ is defined in the same manner. The scaling of the lifetime with system size for $\alpha = 1.5$ is shown in Fig.~\ref{fig:life_time_plot}(d), where the magenta circles denote the numerically extracted lifetimes. The dotted line represents the exponential fitting function $t^{*} \sim e^{2.15(9)L}$, which again confirms that the PDTC lifetime grows exponentially with system size. Overall, these results demonstrate that for both values of $\alpha$, the lifetime of the PDTC phase exhibits robust exponential scaling with system size, highlighting the enhanced stability of time-crystalline order in larger systems. 

{\color{black}The prethermal character of the phase is corroborated at the spectral level. The quasienergy gap-ratio statistics of $U_F$, computed within a fixed sector of the global spin-flip symmetry, follow neither the Poisson distribution characteristic of a many-body-localized regime nor the circular-orthogonal-ensemble statistics of a fully thermalizing Floquet system, consistent with a Floquet operator that remains perturbatively close to the structured e = 0 point where heating is parametrically slow. Likewise, the quasienergy $\Pi$-pairing remains intact across the spectrum for weak imperfection and breaks down only for stronger e. Details of both analyses are presented in Appendix VII A (Figs.~\ref{fig:pi_pair_diff_e} and \ref{fig:fig-level-stat}).}

Here it's customary to mention that while the emergent time-crystal phase and its stability have been discussed in the context of the first kind of Floquet protocol, it has been checked with exemplary test cases that such crystalline phase also arises with the second kind of Floquet protocol (Eq.~(\ref{eq:floquet_operator_2})), which is proposed in the context of quantum battery. This is supported through related discussions in Sec.~(\ref{sec:battery}).

}}

\begin{figure}[t!]
    \centering
    \includegraphics[scale = 0.31]{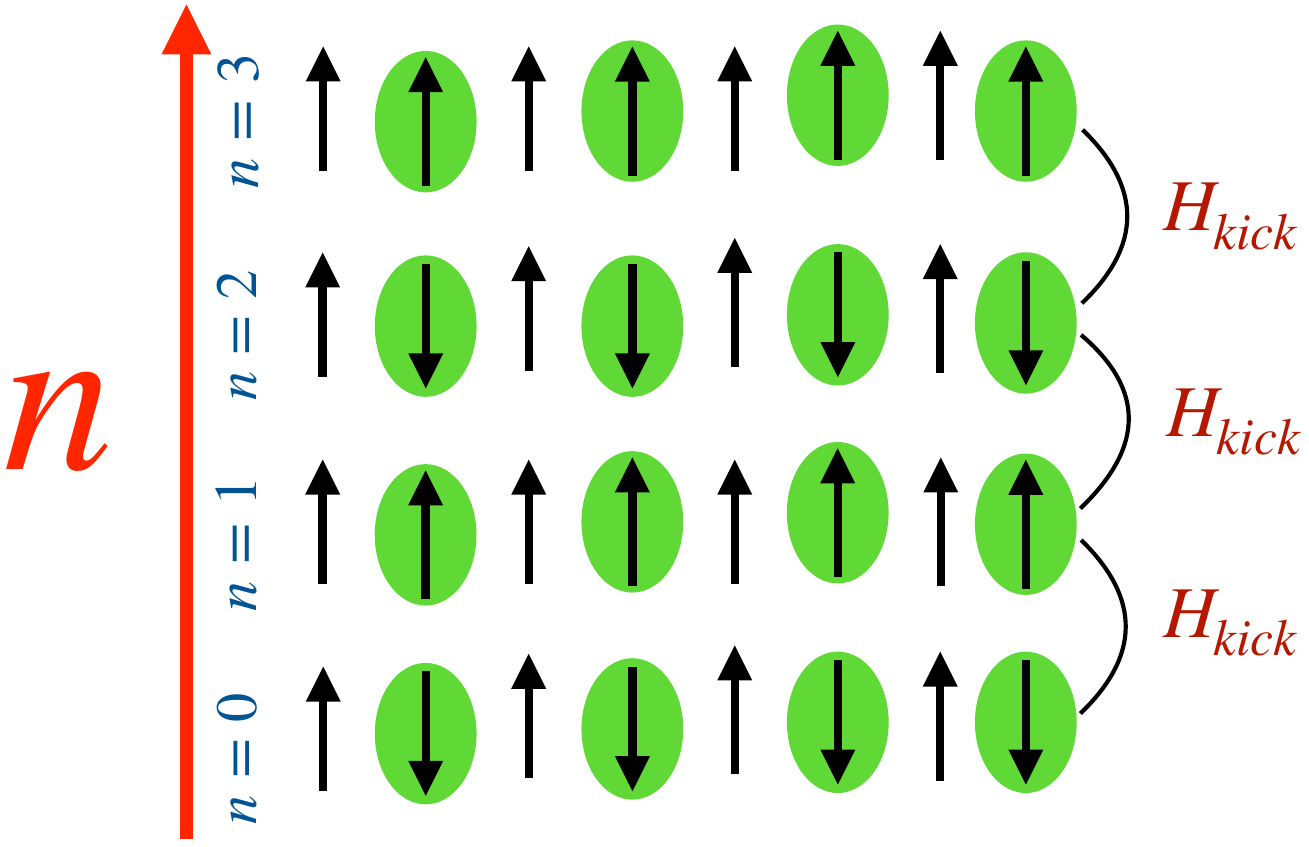}
    \caption{The schematic diagram illustrates the battery charging procedure. The battery initially starts in the ground state, $|x_{\text{min}}\rangle = |\uparrow \downarrow....\uparrow\downarrow\rangle$.  During the evolution of this state, a delta kick acts only on the down-spin positions, highlighted in green color, causing all the down spins to flip to up. As a result, the system reaches its highest excited state, $|x_{\text{max}}\rangle = |\uparrow \uparrow....\uparrow\uparrow\rangle$. This process repeats iteratively, and the battery can store maximum energy after an odd number of stroboscopic time steps. This diagram specifically corresponds to the case where $e=0$.}
    \label{fig:schamatic}
\end{figure} 
\begin{figure*}[t!]
    \centering
    \includegraphics[scale = 0.4]{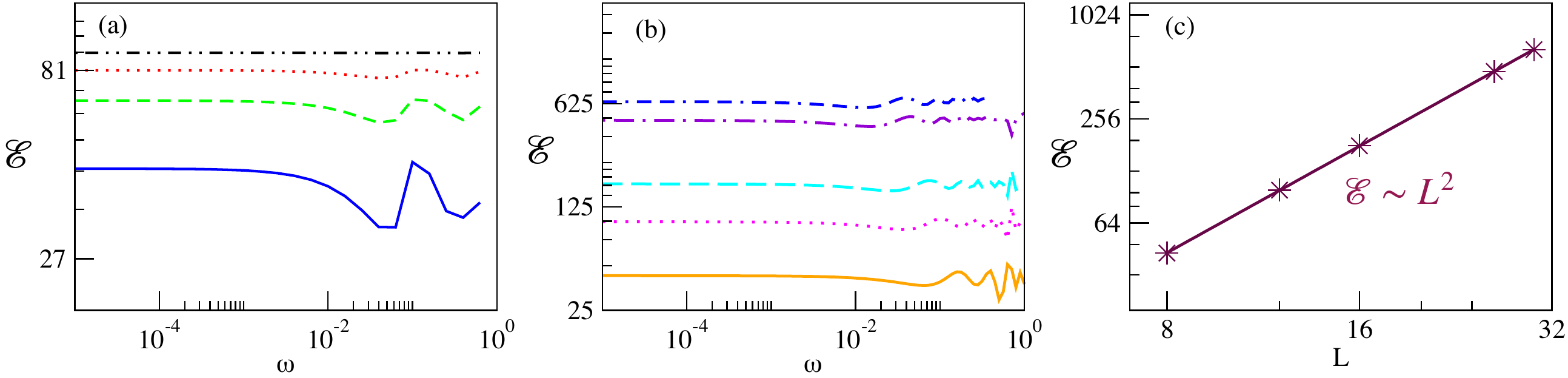}
    \caption{ {\bf (a)} This plot depicts the ergotropy $\mathcal{E}$ of the battery as a function of $\omega$ for different $n = 5$ (black), $31$ (red), $51$ (green), $81$ (blue), by fixing $e=0.01$, $\alpha = 1$ and $L = 10$. The behaviour of $\mathcal{E}$ at the DTC phase (up to $\omega = \omega_c$) is initially constant for a fixed $n$. After $\omega_c$ the $\mathcal{E}$ becomes oscillatory, implying a non-DTC phase. {\bf (b)} This figure presents the $\mathcal{E}$ with respect to $\omega$ for different system size $L = 8$ (orange line), $12$ (magenta dotted line), $16$ (cyan dash-dash line),  $26$ (violet dash-dot line), $30$ (blue dash-dash-dot line) for $n=51$, $e=0.01$ and $\alpha = 1$. From the plots, it is obvious that there is a growth of $\mathcal{E}$ with increasing $L$. We have used the DMRG method to calculate $\mathcal{E}$ for higher system sizes. {\bf (c)}  To extract the scaling of the exponent of $\mathcal{E}$ with $L$, this plot is  presented. The maroon star denotes the numerically extracted data points, and the maroon line is the fitting function showing the scaling relation $\mathcal{E} \sim L^{2.042(2)}$. }
    \label{fig:DE_vs_w}
\end{figure*}

\section{Application as Quantum Battery}
\label{sec:battery}
Quantum batteries are quantum-mechanical systems designed to store and release energy by harnessing intrinsic quantum effects. 
 The well-established charging protocols of the quantum battery have been demonstrated through quench dynamics \cite{Hogan_2024} and Floquet periodic driving \cite{Bukov04032015, Eckardt2017}. In general, the occurrence of a quantum advantage with super-linear scaling in the stored energy or power output in many-body systems of energy charging appears to be rare. Ever since the introduction of Quantum Batteries, many theoretical studies have been conducted to analyze their performance \cite{Alicki2013,Ghosh2022,Juli2020,Ferraro2018, Andolina2019,Santos2019, Rossini2020, Sen2021,Crescente2020, Crescente2022,Ghosh2020b,Ghosh2021c,konar2022b,Bhatta2024, chaki2023, sen2023noisy, shukla2025optimizing, sarkar2025fluctuation, Paulino2025, Mukherjee2024, PhysRevLett.124.170602}. These studies have highlighted significant quantum mechanical phenomena, such as indefinite causal order \cite{Zhu2023} and many-body localization \cite{Rossini2019b, Arjmandi2023}. The recent studies show that by applying floquet dynamics to many-body systems, particularly in the context of an Ising chain, the performance of the battery typically does not yield a quantum advantage \cite{Mondal2022}. However, a recent study suggests that systems with long-range interactions can feasibly achieve quantum advantages, particularly in the form of super-linear scaling of the normalized charging power~\cite{puri2024floquet}.


 In this work, we propose that the PDTC phase realized via Floquet-driven spin chain offers a promising avenue and a natural platform for quantum energy storage. The inherent stability of the PDTC dynamics, characterized by robust period-doubling oscillations persisting over long timescales, ensures that stored energy remains protected against small perturbations. Moreover, the spatially varying interaction profile $j^\alpha$ in our work allows for a tunable platform, which can potentially be exploited to enhance both the storage capacity and the charging power of the battery. By leveraging the non-equilibrium order of the PDTC phase, our approach provides a novel mechanism for realizing stable quantum batteries. 

{\color{black}{Since the couplings $j^{\alpha} > 0$ in $H_B$ have open boundaries, the Hamiltonian hosts two degenerate ground states with perfect antiferromagnetic (Neel)-ordering, $|\uparrow\downarrow\uparrow\downarrow\cdots\rangle$ and $|\downarrow\uparrow\downarrow\uparrow\cdots\rangle$, related by the global spin-flip symmetry. Considering that the system is initiated in one of these two states, say in 
\begin{equation}    |x_{\text{min}}\rangle = |\uparrow \downarrow \uparrow \downarrow \cdots \rangle,
\end{equation}
where we assume $L$ to be even, the  corresponding eigen energy (ground state energy) is given by 
 \begin{equation}
 \label{E_min}
E_{\text{min}} = -\sum_{j=1}^{L-1} j^\alpha.
\end{equation}
On the other hand, the most excited energy label belongs to the configuration of ferromagnetic order in which all spins are aligned in the same direction. By performing spin flipping operation to the alternate spins of the ground state, the highest excited state can be accessed, i.e., 
\begin{equation}
  |x_{\text{max}}\rangle = \prod_{j = 1}^{L/2} \sigma_x^{2j} |x_{\text{min}}\rangle = |\uparrow \uparrow \uparrow \uparrow \cdots \rangle.
\end{equation}
The energy corresponding to this state $|x_{\text{max}}\rangle$ is given by
\begin{equation}
\label{E_max}
E_{\text{max}} = \sum_{j=1}^{L-1} j^\alpha.
\end{equation}

}}

The charging process is implemented by applying a time-dependent field $V(t)$ that couples to the battery Hamiltonian $H_B$. The resulting total Hamiltonian, which we refer to as the charger, is given by  
 \begin{equation}
     H_c = H_B + V(t),
 \end{equation}
where $V(t)$ is the time-dependent interaction. Initially, the system (battery) is prepared in its ground state $|x_{\text{min}}\rangle = |\uparrow \downarrow ....\uparrow\downarrow\rangle$ with energy $E_{\text{min}}$.  Now, by activating the $V(t)$, the battery is set to store energy through the delta kick at a regular interval, $T$. In this charging process, the field $V(t)$ serves as the mechanism to transfer energy to the battery. The maximum possible energy can be stored by driving the system to its highest excited state. This can be achieved via rotation of alternative spins through the unitary operator $H_{kick}=\sum_{j = 1}^{L/2}  \sigma_x^{2j}$. For this choice of $H_{kick}$, the corresponding floquet operator is $ \tilde{U}_F = e^{-i\Phi\sum_{j=1}^{L/2}  \sigma_x^{2j}} e^{-i T H_B}$. All the battery performance analyses are performed using the Floquet operator, $ \tilde{U}_F $. \\

{\color{black}{\emph{Battery performance quantifiers: ergotropy and power.}   The energy stored in the battery after every period $n$ is expressed as
 \begin{align}
    \label{eq:kitaev}
  \Delta E = \text{Tr}[\rho(nT) H_B] - \text{Tr}[\rho_0 H_B],
\end{align}
where $\rho(nT) = \tilde{U}_F^{n} \rho_0 \tilde{U}_F^{n\dagger}$ is the density matrix of the battery after $n$ periods, and the $\rho_0$ is the density matrix corresponding to the initial state of the battery. Although the battery stores an energy $\Delta E$, not all of it can generally be extracted. The maximum extractable energy is referred to as the ergotropy, defined as
\begin{equation}
    \mathcal{E} =  \text{Tr}[\rho(nT) H_B] - \min_{U} \operatorname{Tr}\!\left[ H_B U \rho_0 U^\dagger \right],
\end{equation}
where $U$ denotes a unitary operator. If the density matrix is expressed as $\rho=\sum_k r_k |r_k\rangle\langle r_k|$, with $r_k$ arranged in descending order, the minimization of second term in $\mathcal{E}$ is achieved by the passive state associated with the battery Hamiltonian $H_B$, given by $\rho_{\mathrm p}=\sum_k r_k |E_k\rangle\langle E_k|$, where $|E_k\rangle$ are the eigenstates of $H_B$ ordered by increasing eigenenergy. In our case, the battery is initially prepared in its ground state, which is a pure state. Since unitary evolution preserves purity, the density matrix has a single nonzero eigenvalue $r_1=1$, while all other $r_k$ vanish. Therefore, the passive state reduces to
$\rho_{\mathrm p}=|E_{\min}\rangle\langle E_{\min}|$,
implying that the ergotropy coincides with the stored energy $\Delta E$.

The power output of the battery is defined as
\begin{equation}
    P = \frac{\Delta E}{t},
\end{equation}
where $t=nT$.  To enable a fair comparison between different charging protocols, it is customary to normalize the Hamiltonian such that its energy spectrum remains bounded within a given interval, say $[-1,1]$ \cite{konar2022b}. It also lets one to avoid the trivial situations, such as enhancing the battery power simply by multiplying the battery Hamiltonian by a constant factor greater than unity. Accordingly, the battery Hamiltonian is normalized as
\begin{equation}
\frac{1}{E_{\max} - E_{\min}}
\left[ 2H_B - (E_{\max} + E_{\min}) I \right]
\;\to\; H_B.
\end{equation}
All power calculations presented in this work are performed using the normalized Hamiltonian. \\

\emph{Battery performance.} The primary objective of a quantum battery is to maximize the stored energy and efficiently extract it as useful work. In the present case, the battery is initially prepared in the ground state of the battery Hamiltonian. Since the subsequent Floquet evolution remains unitary, the battery state stays pure throughout the dynamics. Consequently, the ergotropy exactly coincides with the stored energy, $\mathcal{E} = \Delta E$, implying that the stored energy is fully extractable from the battery. The schematic diagram in Fig.~\ref{fig:schamatic} illustrates the mechanism of energy storage in the battery through the action of periodic kicks. Time is represented along the vertical axis, with stroboscopic steps labeled by $n$.  The encircled (even) spins are subjected to the periodic kick. This results in driving the system from the initial state $|x_{\text{min}}\rangle$ ($n=0$) to $|x_{\text{max}}\rangle$ at $n=1$. Upon applying the second kick ($n=2$), the dynamics similarly act on the previously rotated spins (marked in green), inducing another rotation of $\pi/2$. This action flips these spins back to their original down-spin orientation. Consequently, the system returns to its initial product state, $|x_{\text{min}}\rangle$, completing one full cycle. {{\color{black}{This cyclic behavior of spin rotations in this system persists for long times due to the PDTC property}}}. Importantly, at odd stroboscopic times, when all spins are in the up state (corresponding to the most excited state of the battery Hamiltonian), the stored energy of the battery, $H_B$, reaches its maximum value. \\

At odd stroboscopic steps, the system always reaches the fully excited configuration (maximum stored energy), and because of the DTC property, this cycle persists indefinitely without decay. The stability of stored energy, or ergotropy, is due to DTC, and this makes it advantageous compared to conventional charging protocols in Floquet-driven systems, where stored energy rapidly decays with time due to heating. In the following, we investigate the performance of the PDTC battery in the presence of a small perturbation ($e \neq 0$). A desirable quality of a practical quantum battery lies in the stability and robustness of its stored energy against imperfections; it is crucial to test whether the time-crystalline energy storage persists beyond the ideal case. While we majorly perform numerical analysis for the non-ideal cases, it is possible to obtain analytical expressions for the stored energy, which is equivalent to ergotropy, for small system sizes. Here we provide its expression for $L=4$, and odd $n$ in the DTC phase (see Appendix~\ref{appendix_battery} for derivation). The expression of $\mathcal{E}$ is given by
\begin{align}
\label{eq:delta_E}
\mathcal{E} &= \Delta E^{2m-1} \nonumber \\
&= \mathrm{Tr}[\rho H_B] - \mathrm{Tr}[\rho_0 H_B] \nonumber\\
&= \left(E_{\text{max}} - E_{\text{min}}\right) 
   + \frac{e^2\pi^2}{4} \Big[\,1+m^2 \nonumber\\
&\quad + 2m \cos\!\big((E_{\uparrow\uparrow\uparrow\uparrow} 
   + E_{\uparrow\uparrow\uparrow\downarrow})T\big)\,
   E_{\uparrow\downarrow\uparrow\uparrow} \nonumber\\
&\quad + \big(1+m^2 
   + 2m \cos\!\big((E_{\uparrow\uparrow\uparrow\uparrow} 
   + E_{\uparrow\downarrow\uparrow\uparrow})T\big)\big)\,
   E_{\uparrow\uparrow\uparrow\downarrow} \Big],
\end{align}
where $n$ is $2m-1$. Here $ E_{\uparrow\uparrow\uparrow\downarrow} $ and $ E_{\uparrow\downarrow\uparrow\uparrow} $ are the energies correspond to the states $ |\uparrow\uparrow\uparrow\downarrow\rangle $ and $ |\uparrow\downarrow\uparrow\uparrow\rangle $, respectively. For larger system sizes, obtaining a closed-form analytical expression becomes intractable.

For larger system sizes, we compute the ergotropy $\mathcal{E}$ numerically using the density matrix renormalization group (DMRG) within the matrix product state (MPS) formalism. We present $\mathcal{E}$ as a function of $\omega$ for $\alpha = 1$, $e=0.01$ and $L=10$, where $\omega$ quantifies the deviation from  $T$ by considering $T = \pi/2L + \omega$ in the  Fig.~\ref{fig:DE_vs_w}(a). Cases with several odd stroboscopic times $n = 5(\text{black}), 31(\text{red}), 51(\text{green}), 81(\text{blue})$ are demonstrated. The behavior of $\mathcal{E}$ reveals that, within the DTC phase, it remains nearly constant with changing $\omega$ up to a critical deviation $\omega = \omega_c$. Beyond $\omega_c$, the $\Delta E$ exhibits oscillation, indicating the onset of the non-DTC phase. In Fig.~\ref{fig:DE_vs_w}(b), we presents $\mathcal{E}$ as a function of $\omega$ for different system sizes $L = 8$ (orange line), $12$(magenta dotted line), $L =16$ (cyan dash-dash line), $26$ (violet dash-dot), $30$ (blue dot-dash-dot) for $n=51$ and $e = 0.01$. The plot demonstrates that  the storage energy $\Delta E$ is also increases with increasing $L$ throughout all the regions. However, in the DTC phase, the $\Delta E$ maintains stability, and we explicitly focus on this regime. The Fig.~\ref{fig:DE_vs_w}(c) shows that the $\mathcal{E}$ scales with $L$ in the DTC phase as $\mathcal{E} \sim L^{\beta}$, where $\beta = 2.042(2)$ for $\alpha = 1$, suggesting a super-linear scaling of $\mathcal{E}$ with $L$.

We further investigate how the scaling exponent $\beta$ associated with $\mathcal{E}$ ($=\Delta E$) varies with changing the parameter $\alpha$ (see Fig.~\ref{fig:battery_beta_vs_alpha}). The circles are the data points extracted numerically for various $\alpha$. The solid line is the fitting function given by $\beta \sim  \alpha^{1.030(7)} + 1.062(7)$. This fitting suggests a linear relationship of the scaling exponent $\beta$ with respect to $\alpha$, confirming that better quantum advantage can be extracted by tuning $\alpha$ to higher values, as one may expect. \\

\begin{figure}[t!]
    \centering
    \includegraphics[width=0.42\textwidth]{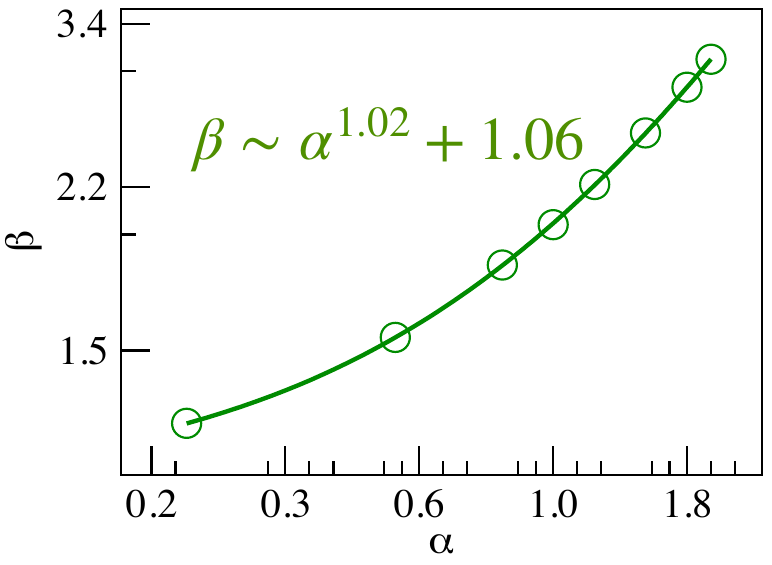}
    \caption{ The scaling exponent of $\Delta E$ with $L$, denoted as $\beta$, is also depends on the parameter $\alpha$. In this plot, the $\beta$ is presented along the y-axis and $\alpha$ is along the x-axis. The green circles are numerically extracted data points of $\beta$ with varying $\alpha$. The green solid line is the fitting function. The $\beta \sim \alpha^{1.030(7)} + 1.062(7)$ is the best-fitted function. This plot reveals that the $\beta$ scales linearly with $\alpha$.}  
    \label{fig:battery_beta_vs_alpha}
\end{figure}

\emph{Power Advantage of the DTC Phase over the Non-DTC Phase.} In Fig.~\ref{fig:power_comparism}, we present a comparative analysis of the normalized charging power of the battery in the DTC and non-DTC phases. The blue circles represent the power as a function of odd Floquet cycles $n$ in the DTC phase ($\omega = 10^{-5}$), while the red squares correspond to the non-DTC phase ($\omega = 10$). Qualitatively, both cases exhibit a decrease in power with increasing $n$. However, the decay of power in the DTC phase is considerably slower compared to the non-DTC phase. As a result, at relatively large $n$, the DTC battery retains a significantly higher charging power than the non-DTC battery.  This demonstrates that the PDTC-assisted battery exhibits enhanced stability and robustness of the charging process over long-time dynamics. In the inset of Fig.~\ref{fig:power_comparism}, we further show the normalized charging power as a function of $n$ for different system sizes $L$ while fixing $\omega = 10^{-5}$, corresponding to the DTC phase. The results reveal that, after normalization of the battery Hamiltonian, the charging power becomes nearly independent of the system size in the DTC phase. Therefore, the apparent superextensive scaling observed before normalization disappears under fair normalization, indicating the absence of a genuine scaling advantage in the charging power.

However, as mentioned before, the real advantage of the DTC phase, however, comes in terms of robustness through emergent subharmonic responses that persist even in the presence of imperfections through initial state preparation and deviations from the perfect driving, and hence is superior when compared to standard time-dependent protocols that require parameter fine-tuning or synchronizing measurement accurately with the desired optimal time during transient dynamics. 

{\color{black}{Although the system is fundamentally a PDTC, its exponential lifetime ($t^*\sim e^{2L}$) is sufficiently long compared to operational battery timescales, enabling stable energy storage and extraction without loss.}} \\

}}

{\color{black}{

\emph{Battery power in open system setting.}   Due to the unavoidable interaction between the battery and the environment, decoherence, dissipation, and energy exchange with the surroundings can occur during the charging process. In this section, we analyze the effect of local dephasing noise on the performance of the DTC battery. We consider the Markovian dephasing dynamics between two consecutive instantaneous kicks or rotations governed by the Lindblad master equation,
\begin{equation}
\label{open_dephasing}
\partial_t \rho = -i[H_B,\rho] + \sum_{\mu}k_{} \left( L_{\mu}\rho L_{\mu}^{\dagger} - \frac{1}{2} \left\{ L_{\mu}^{\dagger}L_{\mu},\rho \right\} \right),
\end{equation}
where $L_{\mu}=\sigma_z$ corresponds to the local dephasing process with dephasing strength $k$. In Fig.~\ref{fig:battery_open_dynamics}, we present the normalized output power as a function of odd Floquet cycles $n$ in the presence of local dephasing for different system sizes $L=4$, $6$, and $8$. Qualitatively, all the curves exhibit a similar behaviour, where the charging power gradually decreases with increasing $n$. However, in the presence of environmental dephasing, the decay of power becomes faster for larger system sizes at long times. In the previous section, corresponding to the closed-system dynamics, we showed that the normalized charging power remains nearly independent of the system size $L$. In contrast, within the open-system framework, as one may expect, the battery performance deteriorates with increasing system size due to the enhanced influence of decoherence and environmental effects.

}}

\begin{figure}[t!]
    \centering
    \includegraphics[scale = 0.3]{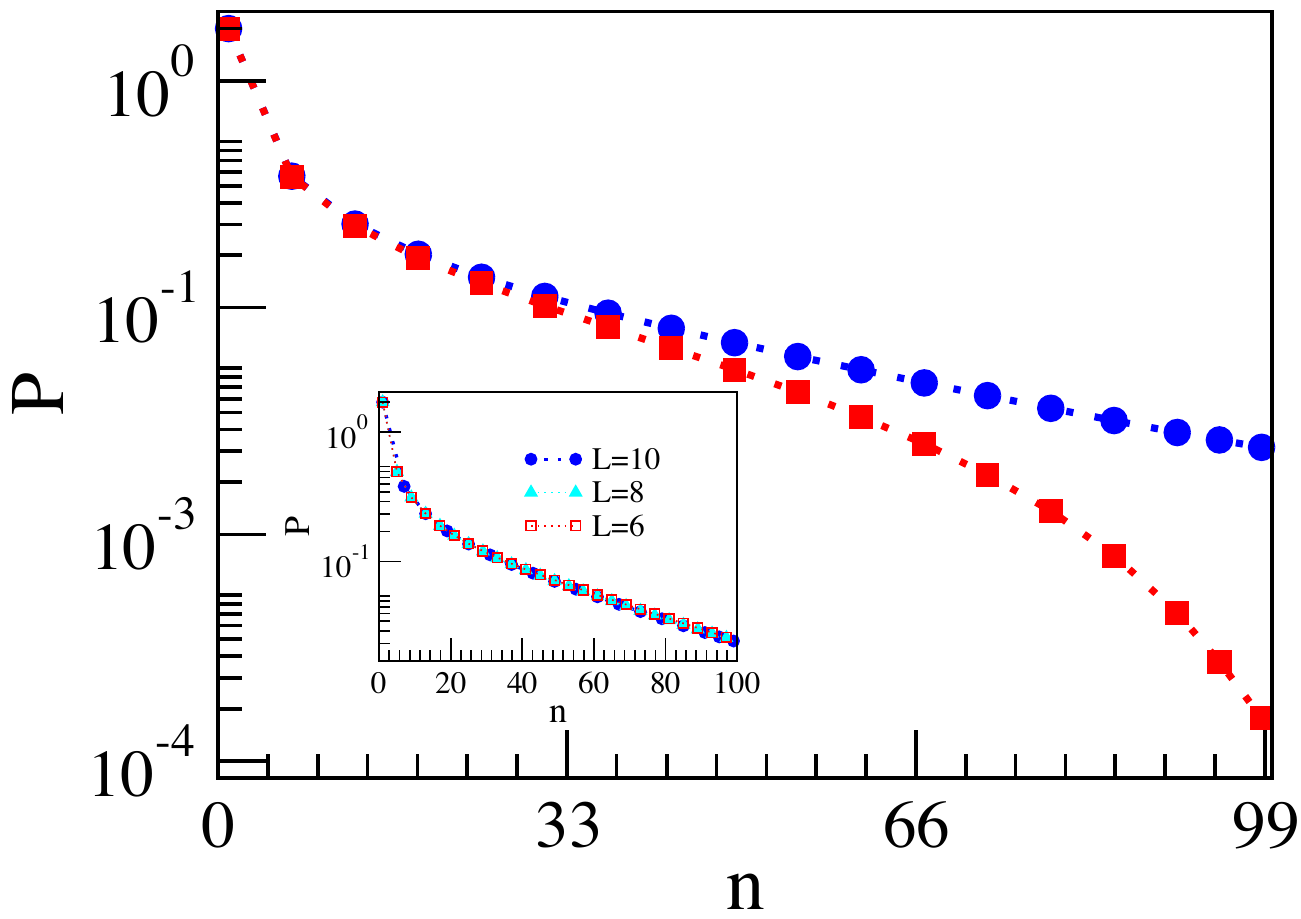}
    \caption{   {\color{black}{  The plot shows the normalized charging power $P$ as a function of odd Floquet cycles $n$ for $L=10$. The blue circles correspond to the DTC regime with $\omega = 10^{-5}$, while the red squares represent the non-DTC regime for fixed $\omega = 10$. The results demonstrate that the battery retains a larger charging power in the DTC phase compared to the non-DTC phase at large $n$. In the inset, we show the normalized charging power for different system sizes $L$ in the DTC regime, demonstrating that the normalized power becomes nearly independent of $L$. All results are presented for $\epsilon=0.01$ and $\alpha=1$.  }}    }
    \label{fig:power_comparism}
\end{figure}

\begin{figure}[t!]
    \centering
    \includegraphics[width=0.38\textwidth]{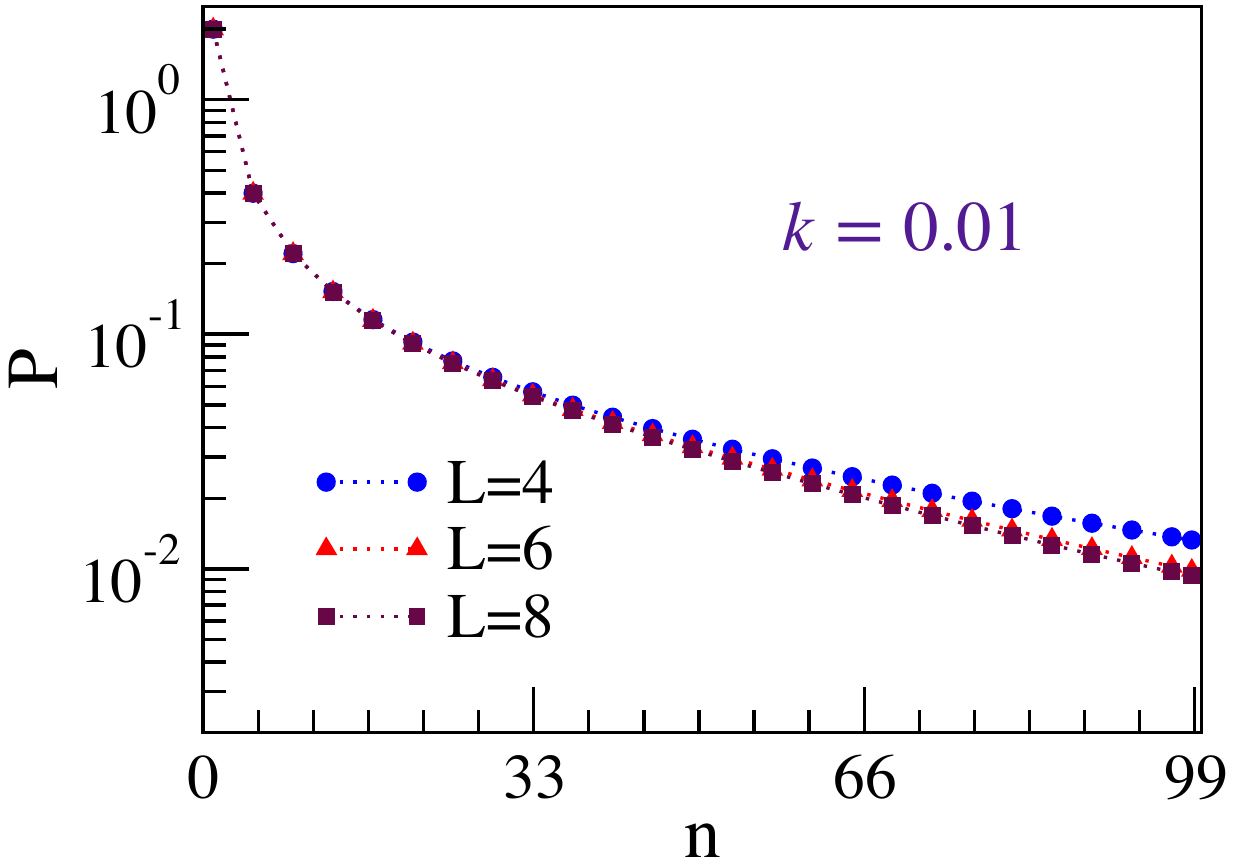}
    \caption{  {\color{black}{The plot depicts the battery power as a function of $n$ in the presence of dephasing dynamics for different system sizes $L=4$, $6$, and $8$ with dephasing strength $k=0.01$ and $e = 0.01$. }}    }  
    \label{fig:battery_open_dynamics}
\end{figure}

\begin{figure*}[t!]
    \centering
    \includegraphics[scale = 0.35]{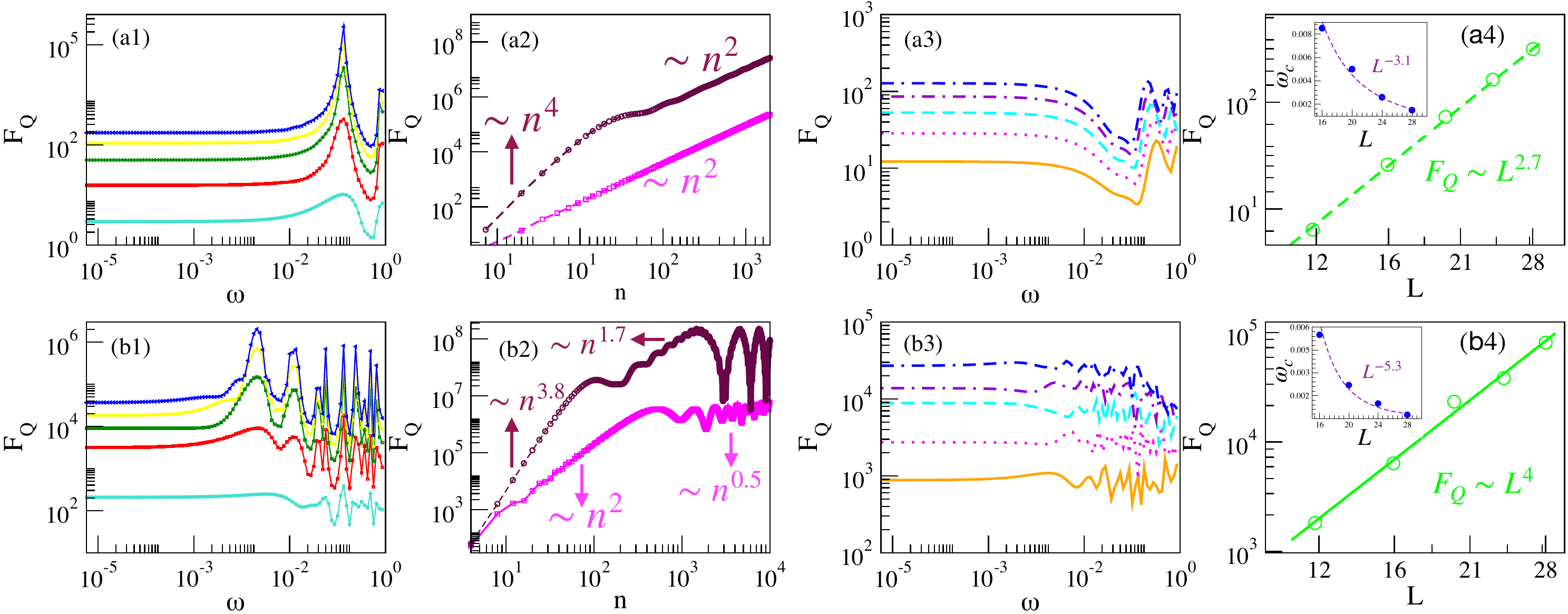}
    \caption{ {\bf For \bm{$\alpha = 0.5$} :} {\bf (a1)} This plot illustrates the behavior of $F_Q$ as a function of $\omega$ at $e=0.01$ and a fix system size $L=8$ with different $n = 4$ (turquosic, circle), $10$ (red, square), $20$ (dark-green, diamond), $30$ (yellow, triangular up), $40$ (blue, triangular left). {\color{black}{Initially, the $F_Q$ shows a plateau over a range of $\omega$ up to $\omega = \omega_c$ (for this particular system size $\omega_c = 0.2$), referring the presence of  DTC phase. Beyond $\omega_c$, $F_Q$ shows a non-trivial oscillating behavior, indicating a non-DTC phase.}} {\bf (a2)} {\color{black}{This figure presents the $F_Q$ versus $n$ for $L=8$.}} The circular dotted line with maroon color represents $F_Q$ at $\omega_c = 0.2$, where the $F_Q$ initially scales as $F_Q \sim n^4$, then it saturates to $F_Q \sim n^2$. The magenta color square with dotted line shows quadratic scaling of $F_Q$ with $n$ when the $\omega$ is across the DTC phase. {\bf (a3)} This figure depicts the $F_Q$ versus $\omega$ for various system sizes $L=12$ (orange line), $16$ (magenta dotted line), $20$ (cyan dash-dash line), $24$ (violet dash-dot), $28$ (blue dash-dash-dot) at $n=4$.  {\bf (a4)} This plot demonstrates the scaling of $F_Q$ with $L$ as $F_Q \sim L^{2.7}$, the green dash-dash line is the fitting function, and the green circle is the numerical data points.  A similar analysis is performed. {\color{black}{In the inset, we show $\omega_c$ versus $L$. The plot reveals that $\omega_c$ decreases with $L$ as $\omega_c \sim L^{-3.1}$.}} {\bf for \bm{$\alpha = 1.5$} :} {\bf (b1)} This plot depicts the $F_Q$ versus $\omega$ for different $n = 4$ (turquosic, circle), $10$ (red, square), $20$ (dark-green, diamond), $30$ (yellow, triangular up), $40$ (blue, triangular left) with $e=0.01$ and $L=8$. {\bf (b2) } {\color{black}{This figure presents scaling of $F_Q$ with $n$. The circular dotted maroon color plot and the magenta color with square line plot present the scaling behaviour of $F_Q$ with $n$ at $\omega=\omega_c$ ($=0.007$ for $L=8$) and at the DTC phase, respectively, for $L=8$.}} {\bf (b3)} This figure depicts the $F_Q$ as a function of $\omega$ for different $L=12$ (orange line), $16$ (magenta dotted line), $20$ (cyan dash-dash line), $24$ (violet dash-dot), $28$ (blue dash-dash-dot) at $n=4$.  {\bf (b4)} The scaling of $F_Q$ with $L$ is presented here, with the green color line as fitting function $F_Q \sim L^4$, and the green circles are numerical data points. {\color{black}{The inset shows the dependence of $\omega_c$ on $L$, with the scaling relation $\omega_c \sim L^{-5.3}$.}}     }
    \label{fig:sensing}
\end{figure*}

\section{Application as Quantum Sensor} 
\label{sec:sensing}
Alongside exploring the DTC phase as a quantum battery, we demonstrate that this system can also function as an enhanced quantum sensor. Quantum sensing aims to estimate physical parameters with exceptional accuracy, and progress has been made by exploiting diverse many-body phenomena, such as criticality near phase transitions \cite{Rams2018,Zanardi2008,Campos2007, Tsang2013, Garbe2020, Chu2021, Montenegro2021b, Mirkhalaf2020, Frerot2018, Zanardi2006, You2007, Zanardi2007c, Ilias2022, Gietka2021, GU2010, mondal2024, agarwal2025}, Floquet dynamics \cite{Mishra2021a, Mishra2022}, scar states \cite{Dooley2021, Desaules2022, Dooley2023b, Guo2023}, and localization-delocalization \cite{He2023, Sahoo2024a,sahoo2024b, Sahoo2025a,debnath2025tilt} (reviews \cite{Montenegro2025,agarwal2025b,Ghosh26}). Building on recent proposals that connect the DTC phase to sensing applications \cite{Liu2023, Montenegro2023, Yousefjani2025,biswas2025,biswas2025floquet}, our work introduces a tunable platform where the grading exponent $\alpha$  governs both the stability of the PDTC phase and the resulting sensing performance. Particularly,  DTC based quantum sensing protocol has been discussed in the presence of Stark Ising interaction, i.e., $\alpha=1$  \cite{Yousefjani2025}. In this work, we generalize the results and show that PDTC-based  sensing remains robust against perturbations and independent of the initial state, while the precision and accuracy can be optimized by tuning $\alpha$. This tunability provides a practical advantage over previous DTC-sensing approaches.

Moreover, quantum sensing has already seen experimental success in atomic clocks \cite{Appel2009, LouchetChauvet2010}, magnetometry \cite{Wasilewski2010, Sewell2012}, interferometry \cite{Mitchell2004, Nagata2007}, and ultracold spectroscopy \cite{Roos2006, Leibfried2004}, underscoring its technological importance. The hallmark property of the DTC phase—spontaneous breaking of discrete time-translation symmetry—ensures sustained coherent dynamics even in the presence of small perturbations. These features position the power-law-graded DTCs as a robust and tunable platform for next-generation quantum sensors.

In our paper, the goal is to estimate the parameter $\omega$, which is introduced by shifting $T$ to $T=\pi/2L+\omega$, and we have chosen $T=\pi/2L$ as at this value of $T$ the system shows the DTC phase. We encode the parameter $\omega$ into a quantum state $\rho(\omega)$. The estimation process involves measuring this state using a set of projection measurement operators $\{\pi_n\}$. Each measurement process the  $n$'th outcome occurs with a probability $p_n(\omega) = \text{Tr}[\rho(\omega)\pi_n]$, which depends on the parameter $\omega$. The precision in estimating the parameter $\omega$ is quantified by the standard deviation $\delta\omega$. According to estimation theory, this uncertainty ($\delta \omega$) is bounded by the Cram\'{e}r-Rao bound $\delta \omega \geq \frac{1}{\sqrt{M F_c}}$, where $M$ denotes the number of measurements, and $F_c$ is the classical Fisher information (CFI) which is defined as
\begin{equation}
    F_c = \sum_n \frac{1}{p_n(\omega)} \left( \frac{d p_n(\omega)}{d\omega} \right)^2,
\end{equation}
which quantifies how sensitively the measurement outcome responds to changes in the parameter $\omega$. The higher value of $F_c$ has a greater ability to measure $\omega$ with excellent accuracy from the measurement results. However, the CFI depends on the specific choice of measurement operators $\{\pi_n\}$. In order to determine the maximum limit of the estimation precision for a given quantum state $\rho(\omega)$, one can maximize the CFI on all possible measurements which will saturate to the quantum Fisher information (QFI), denoted by $F_Q$, which is measurement independent quantity that represents the maximal information that can be extracted about $\omega$. The $F_Q$ can be expressed in terms of the symmetric logarithmic derivative (SLD), $L_\omega$, which satisfies the relation
\begin{equation}
    \partial_\omega \rho(\omega) = \frac{1}{2} \left\{ \rho(\omega), L_\omega \right\},
\end{equation}
where $\{A, B\} = AB + BA$ denotes the anticommutator. The QFI in terms of $L_{\omega}$ becomes $F_Q = \text{Tr}[L_\omega^2 \rho(\omega)]$. The special case when the quantum state is pure, i.e., $\rho(\omega) = |\psi(\omega)\rangle \langle \psi(\omega)|$, the SLD simplifies to $L_\omega = 2 \, \partial_\omega \rho(\omega)$. The QFI then becomes
\begin{equation}
    F_Q = 4 \left( \langle \partial_\omega \psi(\omega)|\partial_\omega \psi(\omega)\rangle - |\langle \psi(\omega)|\partial_\omega \psi(\omega)\rangle|^2 \right).
\end{equation}
This expression tells that $F_Q$ is governed by the rate of state changes for the parameter $\omega$ and the orthogonality between the state and its derivative with respect to $\omega$. In classical sensing, Fisher information can achieve maximum scales linearly with system size $L$, saturating to the standard quantum limit (SQL) $F_Q \sim L$. However, by exploiting quantum features such as entanglement, it is possible to surpass the SQL, and when the $F_Q$ scale quadratically with the system size $F_Q \sim L^2$, which is the so-called Heisenberg limit (HL). {\color{black}{The ultimate precision achievable in our system is determined by the semi-norm of the Hamiltonian $H_B$. The corresponding upper bound on the QFI is given by $F_Q \leq n^2 \|H_B\|_{\mathrm{sn}}^2$, ~\cite{PhysRevLett.98.090401} where the semi-norm is defined as $\|H_B\|_{\mathrm{sn}} = E_{\mathrm{max}} - E_{\mathrm{min}},$ with $E_{\mathrm{max}}$ and $E_{\mathrm{min}}$ given in Eqs.~(\ref{E_max}) and (\ref{E_min}), respectively. In the large $L$ limit, the semi-norm scales as $\|H_B\|_{\mathrm{sn}} \sim L^{1+\alpha}$. Consequently, the upper bound of the QFI takes the form $F_Q \leq n^2 L^{2(1+\alpha)}$. This scaling leads to a maximum achievable QFI.  }}

\begin{figure}[t!]
    \centering
    \includegraphics[width=0.42\textwidth]{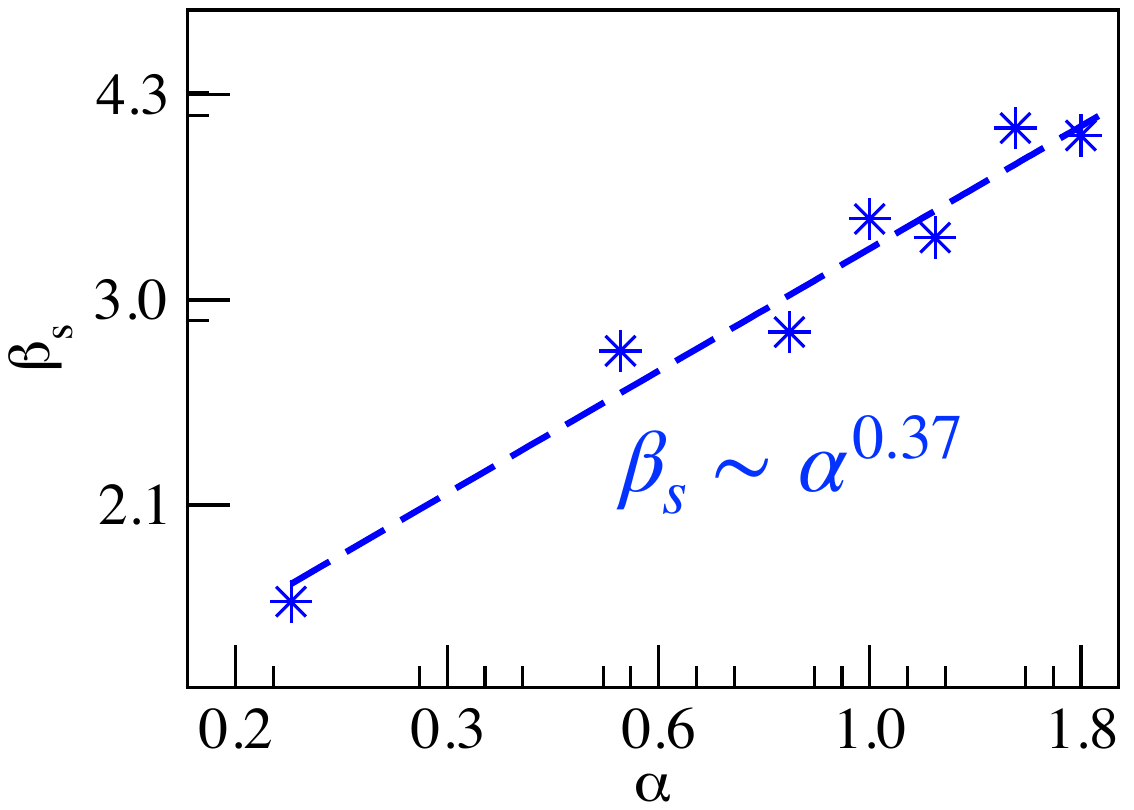}
    \caption{ The scaling exponent of QFI with system-size, $\beta_s$ depends on the value of $\alpha$. This plot presents the $\beta_s$, which is along the y-axis, as a function of $\alpha$ along the x-axis. The blue colored stars are numerically extracted values of $\beta_s$ with varying $\alpha$. The blue dotted line is the fitting function of $\beta_s \sim \alpha^{0.37(3)}$, indicating that $\beta_s$ scales sublinearly with $\alpha$.}  
    \label{fig:sensing_beta_vs_alpha}
 \end{figure} 
{\color{black}The Fig.~\ref{fig:sensing} (a1) depicts the QFI, $F_Q$, as a function of $\omega$ for different stroboscopic times $n = 4$ (turquosic, circle), $10$(red, square), $20$ (dark-green, diamond), $30$ (yellow, triangular up), $40$ (blue, triangular left) for a fixed system size $L=8$ with the perturbation strength $e=0.01$ and $\alpha = 0.5$. This figure demonstrates that $F_Q$ initially remains constant in a small range of $\omega$ for a given $n$, making a plateau; this region is DTC phase. As increasing $\omega$ beyond the plateau, the $F_Q$ first develops a peak and then starts oscillating, indicating the onset of the non-DTC phase. $\omega = \omega_c$ separates these two phases and is named as the transition point. {\color{black}{At $\alpha=0.5$, the transition point is located at $\omega_c \sim 0.2$ for $L=8$. However, $\omega_c$ depends on the system size $L$ and decreases with increasing $L$ as $\omega_c \sim L^{-3.1}$, as shown in the inset of Fig.~\ref{fig:sensing}(a4). This scaling suggests that, in the thermodynamic limit, the transition occurs at an infinitesimal deviation of $\omega$.}} We plot the $F_Q$ as a function of $n$, in the Fig.~\ref{fig:sensing}(a2), for $\omega_c = 0.2$, initially $F_Q$ scales with $n$ as $F_Q \sim n^4$ up to a certain $n$, beyond which the scaling becomes quadratic, $F_Q \sim n^2$ (shown in maroon). However, by fixing $\omega = 0.0001$ in the DTC phase, the scaling of $F_Q$ with $n$ is always quadratic, as shown with a magenta-colored dotted line in Fig.~\ref{fig:sensing} (a2).

The performance of a many-body sensing device is evaluated by the scaling behaviour of $F_Q$ with system size, $L$. Fig.~\ref{fig:sensing} (a3) presents the plots of $F_Q$ as a function of $\omega$ for different system sizes $L=12$ (orange line), $16$ (magenta dotted line), $20$ (cyan dash-dash line), $24$ (violet dash-dot), $28$ (blue dash-dash-dot). These results are carried out for fixed $n=4$, $e=0.01$, and $\alpha = 0.5$. We perform density matrix renormalization group (DMRG) calculations via matrix product state (MPS) formalism for simulating larger system sizes ($L \ge 12$).
The plot reveals that with $L$, the QFI is increasing in DTC ($\omega \le \omega_c$) phase as well as non-DTC phase ($\omega \ge \omega_c$). The system admits proper scaling in the DTC phase, where $F_Q \sim L^{\beta_s}$ with $\beta_s = 2.73(4)$ for $\alpha = 0.5$ (see Fig.~\ref{fig:sensing} (a4)). The QFI in the DTC phase can be described by the following scaling ansatz 
\begin{equation}
    F_Q \sim n^{a} L^{\beta_s},
\end{equation}\
where $a = 2$ and $\beta_s = 2.73(4)$ in the DTC phase for $\alpha = 0.5$. {\color{black}{Hence, the system exhibits quantum advantage in enhanced sensitivity even for $\alpha < 1$}}.

A parallel analysis for another value of $\alpha = 1.5$ is presented in Fig.~\ref{fig:sensing}(b1), which depicts the $F_Q$ with varying $\omega$ for different $n$ for fixed $e=0.01$ and $L=8$. We consider the cases with $n = 4$ (turquosic, circle), $10$(red, square), $20$ (dark-green, diamond), $30$ (yellow, triangular up), $40$ (blue, triangular left).  As observed previously (in the context of $\alpha = 0.5$), here also, the system exhibits DTC and non-DTC phases across the range of $\omega$. {\color{black} {$\omega_c$ turns out to be $0.007$ for $\alpha=1.5$ and $L=8$. For $\alpha = 1.5$, the transition point $\omega_c$ also exhibits a system-size dependence, decreasing with increasing $L$ according to the scaling relation $\omega_c \sim L^{-5.3}$, as shown in the inset of Fig.~\ref{fig:sensing}(b4). }} For fixed $\omega_c = 0.007$, the $F_Q$ with $n$ initially scales for the system size $L=8$ as $F_Q \sim n^{3.8}$. At intermediate values of $n$, however, the scaling exponent decreases, and $F_Q$ instead follows the relation $F_Q \sim n^{1.7}$. For larger value of $n$, $F_Q$ begins to oscillate, as shown in the Fig.~\ref{fig:sensing} (b2) (maroon circle with dotted line). In the DTC phase  ($\omega \le 0.007$), $F_Q$ shows an initial quadratic growth with relation $F_Q \sim n^2$, and at larger times it roughly scales as $F_Q \sim n^{0.5}$ (magenta color square dotted line in the  Fig.~\ref{fig:sensing} (b2)).
Furthermore, Fig.~\ref{fig:sensing}(b3) presents the $F_Q$ as a function of $\omega$ for different system sizes $L$, with fixed parameters $\alpha = 1.5$, $e = 0.01$, and $n=4$. Using DMRG, we compute and plot $F_Q$ for $L=12$ (orange solid line), $16$ (magenta dotted line), $20$ (cyan dashed line), $24$ (violet dash-dotted line), and $28$ (blue dash-dash-dotted line). In the DTC regime, the scaling exponent is extracted as $\beta_s = 4.1(2)$ for $\alpha = 1.5$, as shown in Fig.~\ref{fig:sensing}(b4). Thus, for $\alpha = 1.5$, the scaling ansatz for the QFI in the DTC phase takes the form  $F_Q\sim n^{2} L^{4.1(2)}$.

Finally, we investigate how the scaling exponent $\beta_s$ associated with $F_Q$ depends on the parameter $\alpha$.  The blue stars are the numerically extracted values of the exponent within the DTC phase. The fitting function $\beta_s \sim \alpha^{0.37(3)}$ (blue dotted line) suggests $\beta_s \sim \alpha^{0.37(3)}$, implying a sub-linear scaling of $\beta_s$ with $\alpha$ (Fig.~\ref{fig:sensing_beta_vs_alpha}).
These results highlight that the scaling advantage persists in the estimation of $\omega$ across a range of $\alpha$. \\

 \begin{figure}[t!]
    \centering
    \includegraphics[width=0.5\textwidth]{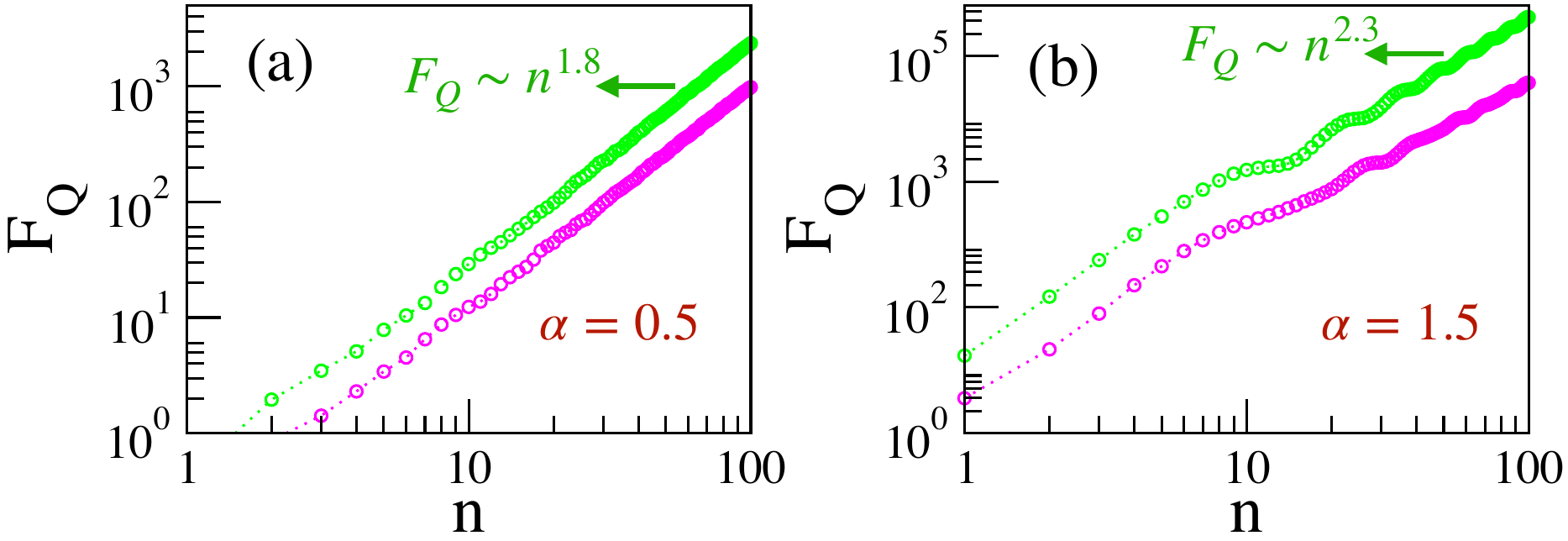}
    \caption{ {\color{black}{(a) The QFI as a function of $n$ is shown for $\alpha = 0.5$ with system sizes $L=6$ (magenta) and $L=8$ (green). The QFI exhibits a scaling behavior $F_Q \sim n^{1.8}$. (b) The variation of QFI with $n$ is presented for $\alpha = 1.5$ with system sizes $L=6$ (magenta) and $L=8$ (green). Here, the QFI scales as $F_Q \sim n^{2.3}$. Both panels correspond to a dephasing rate of $k=0.001$ and $e = 0.01$. } }}  
    \label{fig:open_dynamics_QFI}
 \end{figure}

{\color{black}{

\emph{Sensing in open system setting.} In realistic scenarios, environmental effects are unavoidable. In this panel, we present the effect of Markovian dephasing on our DTC sensor for dephasing rate $k=0.001$ by solving Eq.~(\ref{open_dephasing}). In Fig.~\ref{fig:open_dynamics_QFI}(a), we present the variation of $F_Q$ with $n$ for $\alpha = 0.5$ and for system sizes $L=6$ (magenta) and $L=8$ (green). For both system sizes, the QFI increases with $n$ following the scaling relation $F_Q \sim n^{1.8}$. Moreover, the magnitude of the QFI is larger for higher system sizes, implying an enhanced quantum advantage for larger $L$. A similar analysis is shown for $\alpha = 1.5$ in Fig.~\ref{fig:open_dynamics_QFI}(b), where the scaling of $F_Q$ with $n$ is stronger than that for $\alpha = 0.5$, following $F_Q \sim n^{2.3}$, resulting in a larger quantum enhancement. Here again, we plot the QFI for $L=6$ (magenta) and $L=8$ (green), which reveals that larger system sizes exhibit higher values of QFI.

  }}

\section{Summary}

In summary, we have shown that Floquet-driven spin chains with power-law-graded interactions realize a generalized Stark potential that stabilizes robust PDTC phases. We investigate a range of interaction exponents and demonstrate the appearance of the PDTC phase, making it a flexible and tunable platform for such studies. {\color{black}The cooperative effect of coherent driving and spatially varying couplings renders heating parametrically slow, producing a prethermal window that grows exponentially with system size; identifying the microscopic mechanism behind this stability is an interesting direction for future work.}

{\color{black} A notable aspect of the DTC phase is its natural subharmonic response, in which the system automatically comes back to its original state after a time duration equal to twice the period of the external driving. {\color{black} This characteristic persists throughout the prethermal window, $t \le t^* \sim e^{cL}$, without decay, signaling a discrete time crystal.} This time sequence is a defining feature of the DTC or PDTC phase and acts as an essential mechanism for preserving coherence in non-equilibrium quantum states.  By utilizing this characteristic, we have suggested its use  in the realm of quantum batteries, demonstrating that it can serve as a significant application for energy storage. We show that energy can be effectively extracted from the system's most excited state and the storage energy displays super-linear growth with the system size, with the relation. Additionally, we utilize the DMRG technique to confirm that the super-linear scaling behavior of energy storage also holds for larger system sizes in the presence of small perturbations. The advantage of the present protocol lies not in enhanced charging power but in the robustness of the charging process. The time-crystalline dynamics ensure that the fully charged state reappears periodically and remains stable against small perturbations, in contrast to conventional quench-based charging schemes that require precise timing.} Apart from the energy storage, the time-crystal also offers quantum advantages as a quantum sensor. We have shown that our system can be used as an advanced quantum sensor that can beat HL, a crucial benchmark in quantum metrology. 

In principle, power-law-graded interaction provides a tunable platform for realizing robust PDTC phases that support practical applications, such as designing robust quantum batteries and quantum-enhanced sensing devices. Importantly, the required ingredients, spatially modulated interactions with power-law grading and Floquet driving, are available in current experimental platforms, including trapped-ion chains and Rydberg atom arrays. This makes the proposed mechanism a realistic pathway toward harnessing disorder-free time-crystalline phases for scalable quantum technologies. {\color{black}{A natural direction for future work is to extend the present analysis within an open quantum system framework, allowing one to incorporate decoherence effects in realistic environments and evaluate the performance of the proposed quantum devices.}}

}

\section{Appendix}

In this Appendix, we provide additional analytical and
numerical results to support the findings presented in the main
text. First, we investigate the effects of driving imperfections
on the stability of the PDTC, including quasienergy $\pi$
pairing and level-spacing statistics. Next, we present an analytical treatment of the quantum battery for small system
sizes and examine the robustness of energy storage against
driving imperfections. Finally, we explore the influence of
initial states and driving imperfections on the QFI.
\begin{figure*}[t!]
    \centering
    \includegraphics[scale = 0.5]{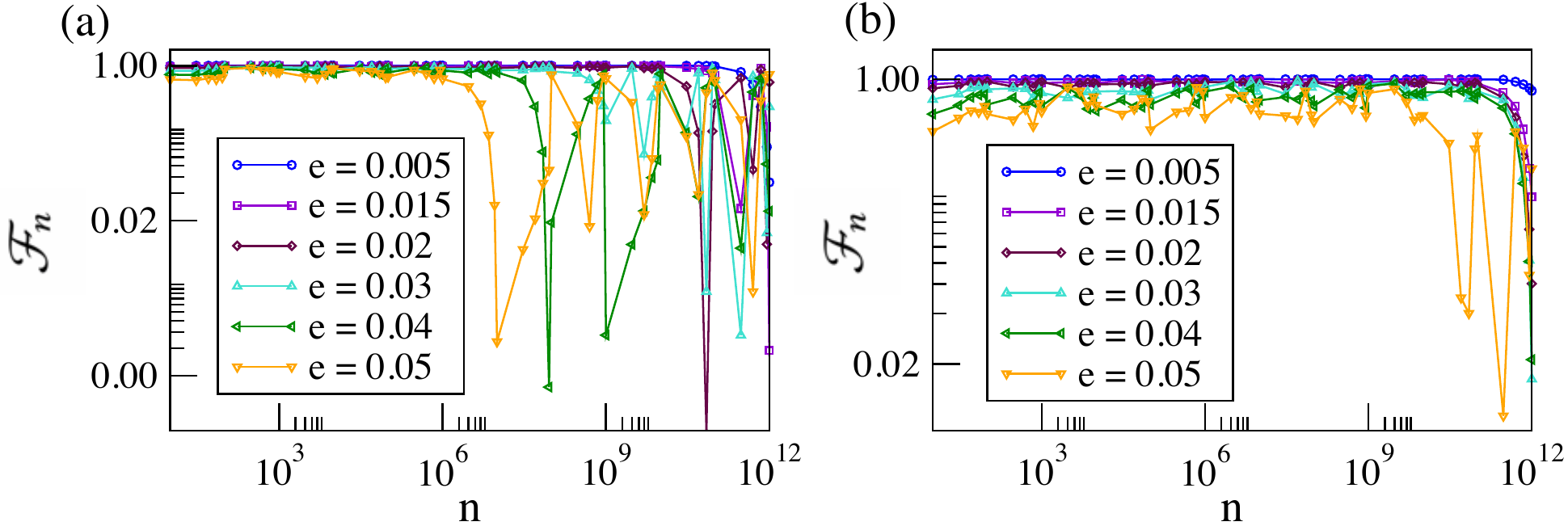}
    \caption{{\color{black}{{\bf(a)}The fidelity $\mathcal{F}_n$ as a function of the stroboscopic time $n$ is shown for different values of $e$ $\in \{0.005, 0.015, 0.02, 0.03, 0.04, 0.05\}$, with fixed $\alpha = 0.5$ and $L=10$. The system is initialized in the state $|\uparrow\downarrow\uparrow\downarrow\dots\rangle$.
  {\bf (b)} A parallel analysis of $\mathcal{F}_n$ versus $n$ is presented for $\alpha = 1.5$, keeping the other parameters the same as in panel (a).} }}
    \label{fig:F_n_vs_n_diff_e}
\end{figure*}

\begin{figure*}[t!]
    \centering
    \includegraphics[scale = 0.35]{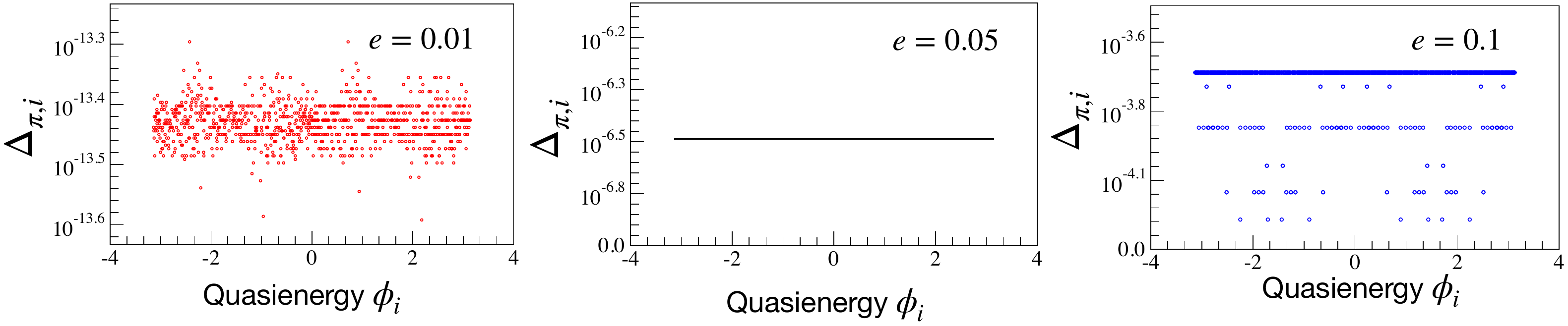}
    \caption{{\color{black}{ The plots show the $\pi$-pairing quasienergy gap, $\Delta_{\pi,i}$, as a function of the quasienergy $\phi_i$ for $\alpha=0.5$ and $L=10$. Panels (a), (b), and (c) correspond to different values of the kick imperfection $e$, respectively.} }}
    \label{fig:pi_pair_diff_e}
\end{figure*}

\begin{figure*}[t!]
    \centering
    \includegraphics[scale = 0.38]{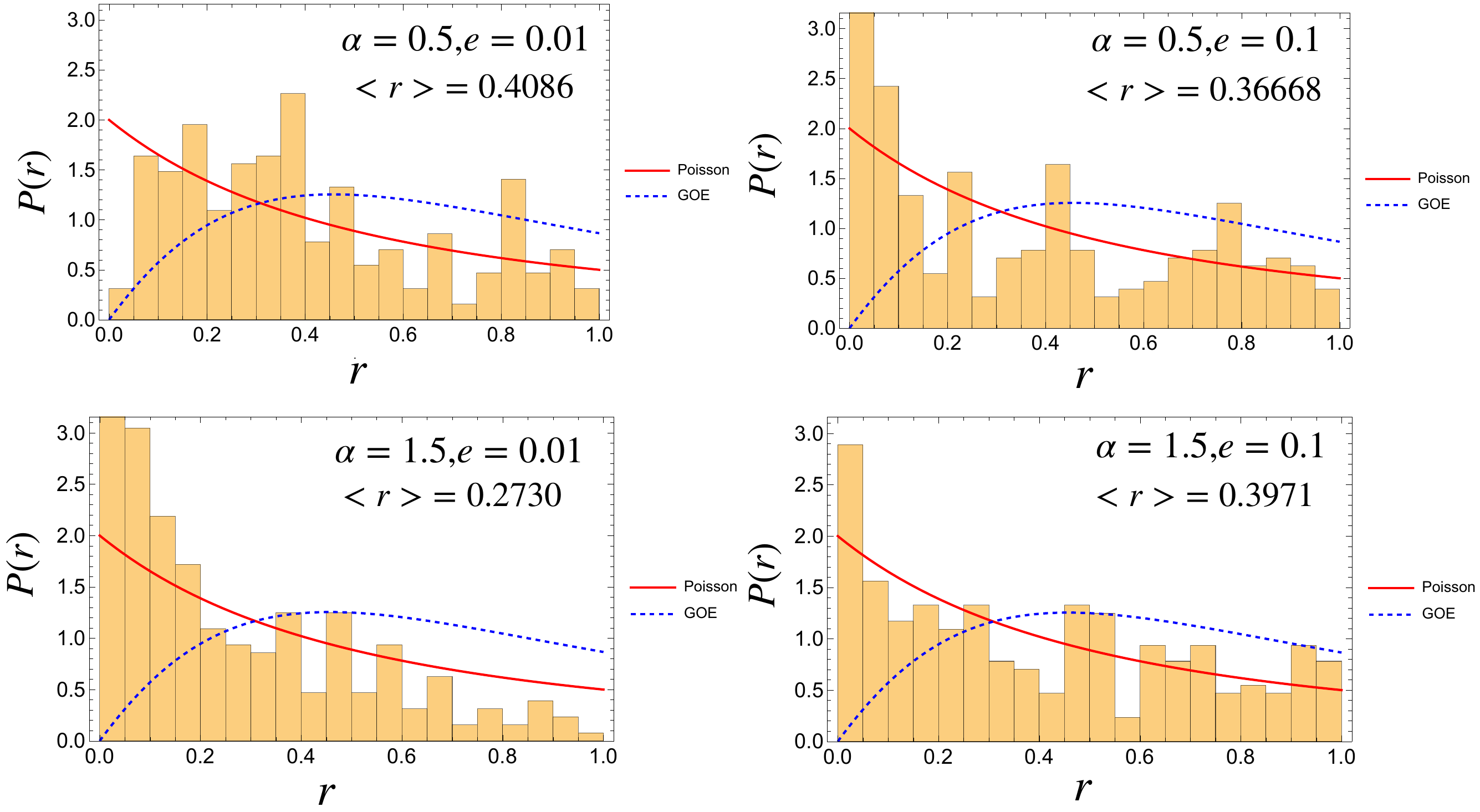}
    \caption{{\color{black}{ Distribution $P(r)$ of the adjacent quasienergy gap ratio of $U_F$, computed within a fixed sector of the global spin-flip symmetry for $L = 10$, at (a) $\alpha = 0.5, e = 0.01$; (b) $\alpha = 0.5, e = 0.1$; (c) $\alpha  = 1.5, e = 0.01$; (d) $\alpha = 1.5, e = 0.1$, with the sector-resolved $\langle r \rangle$ indicated in each panel. Solid and dashed curves: Poisson and Wigner–Dyson references.} }}
    \label{fig:fig-level-stat}
\end{figure*}

\begin{figure}[t!]
    \centering
    \includegraphics[scale = 0.65]{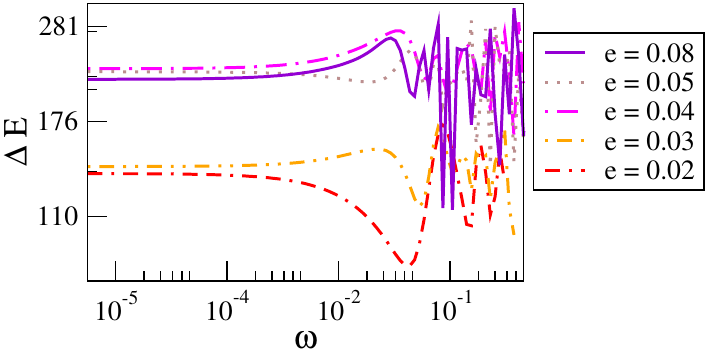}
    \caption{{\color{black}{ The storage energy $\Delta E$ as a function of $\omega$ is shown for $\alpha = 1$ and $L = 20$, considering different values of $e \in \{0.02, 0.03, 0.04, 0.05, 0.08\}$. The initial state is taken to be the ground state of the Hamiltonian.}
   }}
    \label{fig:battery_different_e}
\end{figure}


\begin{figure}[t!]
    \centering
    \includegraphics[scale = 0.65]{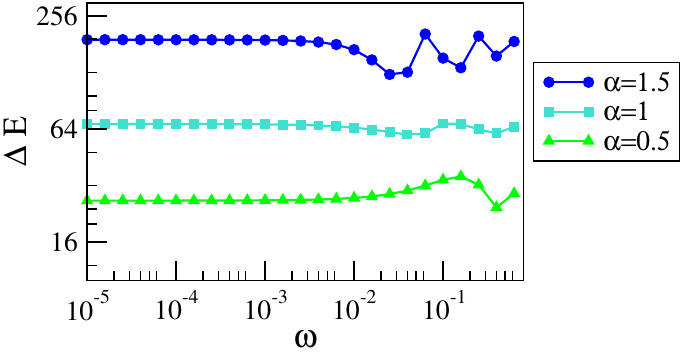}
    \caption{{\color{black}{ The stored energy $\Delta E$ (equivalently, the ergotropy $\mathcal{E}$) as a function of $\omega$ is shown for different values of $\alpha = 0.5$ (green), $1.0$ (turquoise), and $1.5$ (blue) with system size $L=10$ and pulse imperfection $e=0.01$. The battery is initially prepared in the ground state of the Hamiltonian.  }
   }}
    \label{fig:battery_different_alpha}
\end{figure}

\subsection{{\color{black}PDTC: Effects of imperfection, $\Pi$-pairs and level Statistics}}
\label{appendix_timecrystal}

We consider a 1D spin system of size $L$ with initial state
$|x\rangle = |\uparrow\downarrow\uparrow\downarrow\rangle$. In the presence of a perturbation $e$ in the delta kick, the Floquet unitary $U_F$, applied to this initial state, after one period of evolution the resulting state will become

\begin{align*}
    U_F |x\rangle &= (-i)^{L}e^{-E_x T} \sum_{n_f=0}^{L} \Big(cos(\phi)^{L-n_f} (isin(\phi))^{n_f}\Big) \\
    &\Gamma (n_f) |-x\rangle.  \nonumber  \\
\end{align*}

Here $E_x$ is the energy corresponding to the initial state $|x\rangle$,
$\phi = e \pi/2$,
$|-x\rangle$ denotes the all spin-flipped configuration of $|x\rangle$,
$\Gamma(n_f)$ is an operator that generates a linear combination of all states corresponding to $n_f$ spin flips applied to $|-x\rangle$.

For example,
$$
\Gamma(1) |\uparrow\downarrow\uparrow\downarrow\rangle = 
|\downarrow\downarrow\uparrow\downarrow\rangle +
|\uparrow\uparrow\uparrow\downarrow\rangle +
|\uparrow\downarrow\downarrow\downarrow\rangle +
|\uparrow\downarrow\uparrow\uparrow\rangle.
$$

Now, the return amplitude after two Floquet periods is given by
\begin{align}
\langle x | U_F^{2} | x \rangle &=  \nonumber \\
&(-i)^{2L}e^{-2E_x T} \sum_{n_f=0}^{L} \Big( (cos(\phi))^{2(L-n_f)} (isin(\phi))^{2 n_f}\Big) \nonumber  \\
   & \sum_{\{n_f\}}e^{-iE_{-x}\{n_f\} T}, \nonumber
\end{align}
where the sum $\sum_{\{n_f\}}$ runs over all spin configurations generated by $\Gamma(n_f)$, and $E_{-x}\{n_f\}$ is the energy corresponding to the state of all spin configurations resulting from spin flips on $|-x\rangle$. 

{\color{black}{\emph{Imperfection Effect of DTC.}  In the main text, we demonstrated the stability of our DTC for large $L$ by showing that its lifetime increases exponentially with the system size $L$ for different values of $\alpha$. The behavior of the DTC in the presence of imperfection $e$ is presented in Fig.~\ref{fig:F_n_vs_n_diff_e}. For $\alpha = 0.5$ and $\alpha = 1.5$, we plot $\mathcal{F}_n$ as a function of $n$ for different values of $e \in \{0.005, 0.015, 0.02, 0.03, 0.04, 0.05\}$, as shown in Fig.~\ref{fig:F_n_vs_n_diff_e}(a) and (b), respectively. Both figures indicate that smaller values of $e$ correspond to longer lifetimes, whereas increasing $e$ leads to a gradual reduction in the lifetime. However, Fig.~\ref{fig:F_n_vs_n_diff_e}(a) shows that for $\alpha = 0.5$, the DTC remains robust up to large $n$ (approximately $n \sim 10^6$) over a broad range of $e$. Similarly, for $\alpha = 1.5$, the DTC also exhibits robustness up to large $n$ across a wide range of imperfection strengths, as illustrated in Fig.~\ref{fig:F_n_vs_n_diff_e}(b). \\

{\color{black}{ We also examine the stability of the $\pi$-pairing under driving imperfections, $e = 0.01$, $0.05$ and $0.1$. For weak perturbations ($e=0.01$ and $0.05$), the $\pi$-pairing deviation $\Delta_{\pi,i}$ remains very small across the quasienergy spectrum, indicating robust $\pi$-pairing. For larger perturbations ($e=0.1$), the observed deviation, $\Delta_{\pi,i} = \min_{j \neq i}|\phi_j - \phi_i -\pi|_{\text{mod} 2\pi}$ is $O(10^{-4})$, and is yet smaller than the mean level spacing that we define as $2\pi/2^L$. This indicates that a gradual breakdown of the $\pi$-pair structure (Fig.~\ref{fig:pi_pair_diff_e}). Even the largest splittings at the breakdown are more than an order of magnitude below the mean spacing, the identified $\Pi$-partners are genuine spectral pairs. The steep growth of the splitting with increasing $e$ is expected, as high-order perturbative effects are non-negligible.}

The plots indicate that the period-doubled spectral structure remains stable over a broad prethermal regime but eventually breaks down as the perturbation increases. }}

{\color{black}\emph{Level Statistics}: we analyze the quasi-energy level-spacing statistics of the Floquet operator using the adjacent-gap ratio $r_n =
\frac{\min\left(\delta_n,\delta_{n+1}\right)}
{\max\left(\delta_n,\delta_{n+1}\right)}$,
where $\delta_n=\phi_{n+1}-\phi_n$ is the spacing between consecutive quasienergies $\phi_n$. The $\langle r\rangle$ is the mean of the distribution, defined as $\langle r\rangle = \frac{1}{N_r}\sum_{n=1}^{N_r} r_n$, where $N_r$ is the total number of adjacent-gap ratios. The resulting distribution is sub-Poissonian. It deviates from both the Poisson and Gaussian or circular orthogonal ensemble \cite{DAlessio2014LongTime}, indicating that the system does not exhibit the level-spacing statistics characteristic of either a conventional MBL regime or a thermal chaotic regime.  We display corresponding results for $\alpha=0.5$ and $\alpha=1.5$, for different $e$ (Fig.~\ref{fig:fig-level-stat}). The presented results are obtained within a fixed sector of the global spin-flip symmetry $F = \Pi_{j=1}^{L} \sigma_x^j$,  which commutes with both the Ising part and the kick of the Floquet operator, $[F,H_B]=0$ and $[F, e^{-i\Phi\Sigma_j \sigma_x^j}] = 0$. Thus $F$ commutes with $U_F$ itself for arbitrary $e$. Mixing sectors superposes independent spectra and that may cause biases in $\langle r \rangle$ \cite{Giraud2022Probing}. So, we resolve $\mathbb{Z}_2$ symmetry for computing gap-ratio statistics. We confine our analysis in the symmetry sector corresponding to $F=-1$ in order to get rid of any  symmetry-introduced artifact. The corresponding values of $\langle r \rangle$ differ from the Poissonian value $\langle r \rangle_p=0.386$ \cite{Atas2013Distribution}.

Thus, the quasienergy level statistics do not provide evidence for a conventional MBL regime. Together with the eventual Floquet heating observed in the finite-size dynamics, these results suggests emergence of a prethermal DTC rather than an MBL-stabilized DTC.}

}}

\begin{figure*}[t!]
    \centering
    \includegraphics[scale = 0.55]{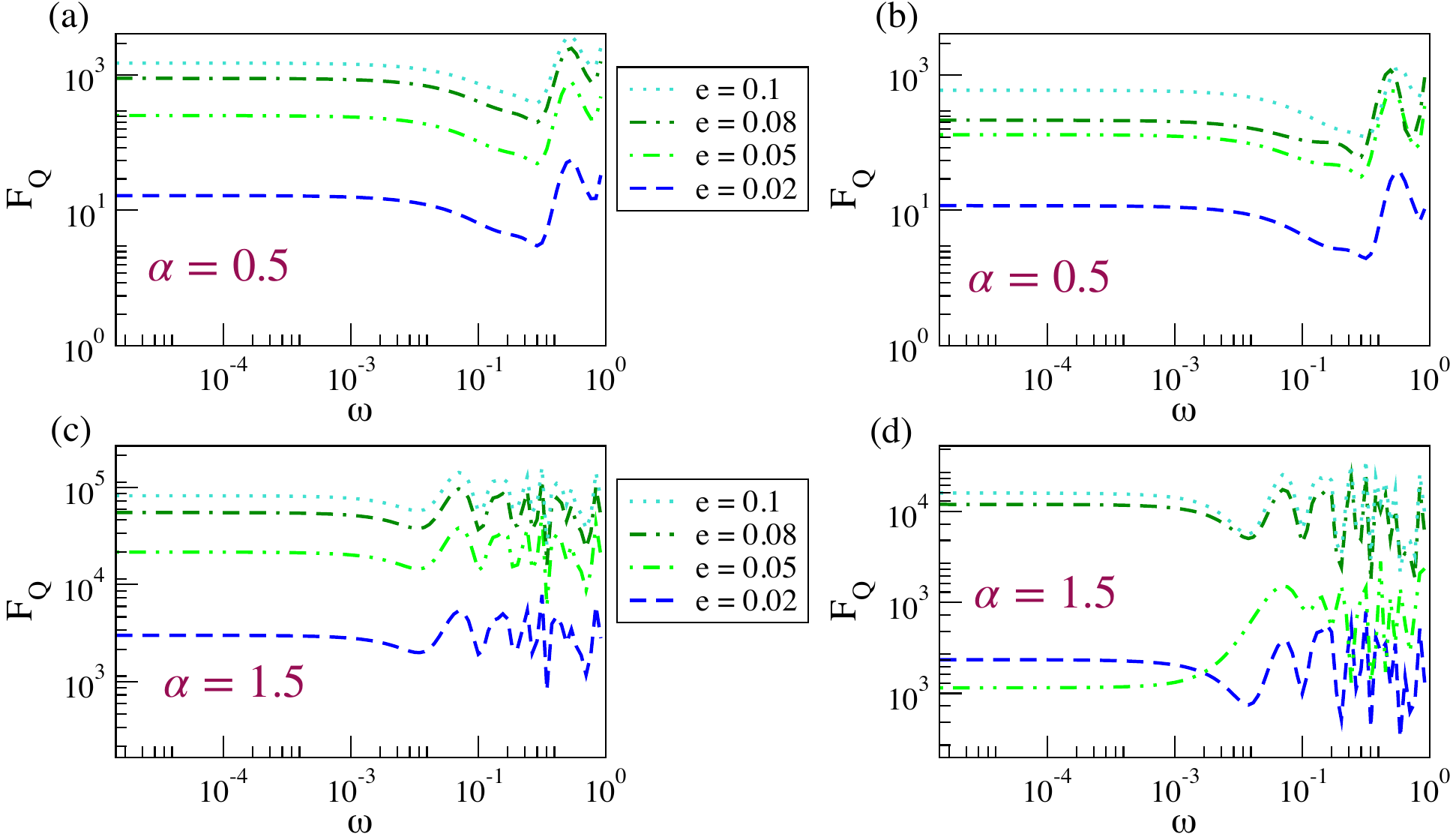}
    \caption{ {\color{black}{{\bf (a)} The QFI with respect to $\omega$ is evaluated for 
$e \in \{0.02, 0.05, 0.08, 0.1\}$, with $\alpha = 0.5$, $L = 10$,  and $n = 4$, using the initial state 
$|\uparrow\downarrow\uparrow\downarrow\ldots\rangle$.  {\bf (b)} The corresponding plot is shown for a randomly chosen initial state under the same parameter regime as in (a). {\bf (c)}  A parallel analysis is carried out for $\alpha = 1.5$, 
with $e \in \{0.02, 0.05, 0.08, 0.1\}$, $L = 10$, and $n = 4$,  where the initial state is again taken as 
$|\uparrow\downarrow\uparrow\downarrow\ldots\rangle$ {\bf (d)} The plot for a randomly chosen initial state under the same parameter settings as in (c) is presented.  } }}
    \label{fig:sensing_diff_initial_st}
\end{figure*}

\subsection{Battery: Small-system analysis}
\label{appendix_battery}

The ground state of the battery Hamiltonian $H_B$ is  
$|x_{\text{min}}\rangle = |\uparrow \downarrow \uparrow \downarrow \cdots\rangle$, corresponding 
The ground state energy is
\[
\langle x_{\text{min}}| H_B |x_{\text{min}}\rangle
= -1^{\alpha} - 2^{\alpha} - 3^{\alpha} - \cdots - (L-1)^{\alpha}
= -\sum_{j=1}^{L-1} j^{\alpha}.
\]

The highest energy state is the fully polarized state $|x_{\text{max}}\rangle = |\uparrow \uparrow \cdots \uparrow \uparrow\rangle$, which is obtained by applying
\[
|x_{\text{max}}\rangle = \prod_{j=1}^{L/2} \sigma_x^{2j} |x_{\text{min}}\rangle,
\]
corresponding to the highest energy of the system is
\[
E_{\text{max}} = \sum_{j=1}^{L-1} j^{\alpha}.
\]

To maximally store the energy from the system, we drive the system from the ground state to the highest excited state by a kicking Hamiltonian through $V(t)$. The kick Hamiltonian is 
\[
H_{\text{kick}} = \sum_{j=1}^{L/2} \sigma_x^{2j}.
\]

We measure the energy storage of $H_B$ at odd multiples of the driving period ($n$). Due to the discrete time-crystalline behavior exhibited by the system with period doubling, the system returns to its initial state after an even number of periods. At these even $n$, the extractable energy from the battery is nearly zero, but at every odd value of $n$, as the system goes to near the highest energy level, the energy storage will be maximum. \\

\begin{figure*}[t!]
    \centering
    \includegraphics[scale = 0.8]{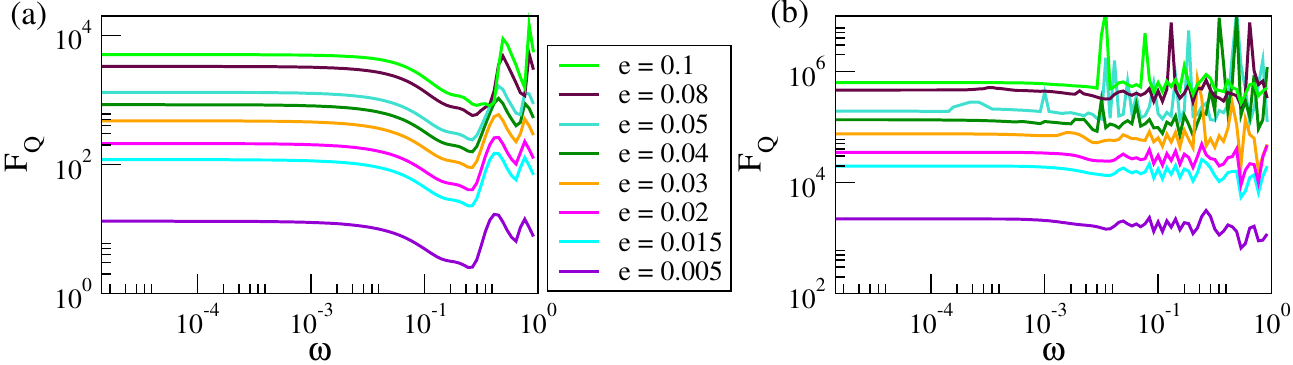}
    \caption{ {\color{black}{{\bf (a)} The QFI with respect to $\omega$ is evaluated for different values of $e$, with $\alpha = 0.5$, $L = 20$, and $n = 4$, using the DMRG method. {\bf (b)}  A parallel analysis is performed for $\alpha = 1.5$ under the same parameter regime as in (a). In both cases, the initial state is taken as $|\uparrow\downarrow\uparrow\downarrow\ldots\rangle$. } }}
    \label{fig:sensing_diff_e}
\end{figure*}

The initial energy of the battery Hamiltonian $H_B$ is given by  
\[
\text{Tr}[\rho_0 H_B] = E_{\text{min}}.
\]  
The energy stored in the battery at time $nT$ is defined as  
\[
E(nT) = \text{Tr}[\rho H_B] - \text{Tr}[\rho_0 H_B],
\]  
where $\rho = \tilde{U}_F^n \rho_0 \tilde{U}_F^{n\dagger}$ is the evolved state after $n$ stroboscope time. In the presence of a small perturbation parameter $e$, the Floquet operator $\tilde{U}_F$ can be written as
\begin{align*}
   \tilde{U}_F &= e^{-i\frac{\pi}{2}(1 - e)\sum_{j=1}^{L/2} \sigma_x^{2j}} e^{-i T H_B} \\
       &= e^{i e \frac{\pi}{2} \sum_{j=1}^{L/2} \sigma_x^{2j}} e^{-i \frac{\pi}{2} \sum_{j=1}^{L/2} \sigma_x^{2j}} e^{-i T H_B} \\
       &= e^{i e \frac{\pi}{2} \sum_{j=1}^{L/2} \sigma_x^{2j}} \tilde{U}_F^0,
\end{align*}
where $\tilde{U}_F^0$ is the Floquet operator for $e=0$.

Expanding the exponential in powers of $e$ via Taylor series and neglecting higher-order terms ($\mathcal{O}(e^2)$), we get
\begin{align*}
   \tilde{U}_F &= \left(I + i e \frac{\pi}{2} \sum_{j=1}^{L/2} \sigma_x^{2j} + \mathcal{O}(e^2)\right) \tilde{U}_F^0 \\
       &\approx \left(I + i e \frac{\pi}{2} \sum_{j=1}^{L/2} \sigma_x^{2j} \right) \tilde{U}_F^0.
\end{align*}

When $e = 0$, the evolution of the ground state under $\tilde{U}_F^0$ is
\begin{align*}
   \tilde{U}_F^0 |x_{\text{min}}\rangle &= e^{-i \frac{\pi}{2} \sum_{j=1}^{L/2} \sigma_x^{2j}} e^{-i T H_B} |x_{\text{min}}\rangle \\
   &= e^{-i T E_{\text{min}}} \prod_{j=1}^{L/2} (-i) \sigma_x^{2j} |x_{\text{min}}\rangle \\
   &= e^{-i T E_{\text{min}}} (-i)^{L/2} |x_{\text{max}}\rangle.
\end{align*}

Now, applying $\tilde{U}_F$ to the ground state $|x_{\text{min}}\rangle$, we obtain
\begin{align*}
   \tilde{U}_F |x_{\text{min}}\rangle 
   &= \left(I + i e \frac{\pi}{2} \sum_{j=1}^{L/2} \sigma_x^{2j}\right) \tilde{U}_F^0 |x_{\text{min}}\rangle \\
   &= e^{-i T E_{\text{min}}} (-i)^{L/2} \times \\
   & \left( |x_{\text{max}}\rangle 
   + i e \frac{\pi}{2} \big( |\uparrow\downarrow\uparrow\uparrow\rangle + |\uparrow\uparrow\uparrow\downarrow\rangle \big) \right).
\end{align*}

Similarly, acting $\tilde{U}_F$ on $|x_{\text{max}}\rangle$ gives
\begin{align*}
   \tilde{U}_F |x_{\text{max}}\rangle 
   &= e^{-i T E_{\text{max}}} (-i)^{L/2} \times \\
   &\left( |x_{\text{min}}\rangle 
   + i e \frac{\pi}{2} \big( |\uparrow\downarrow\uparrow\uparrow\rangle + |\uparrow\uparrow\uparrow\downarrow\rangle \big) \right).
\end{align*}

For the intermediate states, it has form of 
\begin{align*}
   \tilde{U}_F |\uparrow\downarrow\uparrow\uparrow\rangle 
   &= e^{-i T E_{\uparrow\downarrow\uparrow\uparrow}} (-i)^{L/2} \times \\
   &\left( |\uparrow\uparrow\uparrow\downarrow\rangle 
   + i e \frac{\pi}{2} ( |x_{\text{min}}\rangle + |x_{\text{max}}\rangle ) \right), \\
   \tilde{U}_F |\uparrow\uparrow\uparrow\downarrow\rangle 
   &= e^{-i T E_{\uparrow\uparrow\uparrow\downarrow}} (-i)^{L/2} \times \\
   &\left( |\uparrow\downarrow\uparrow\uparrow\rangle 
   + i e \frac{\pi}{2} ( |x_{\text{min}}\rangle + |x_{\text{max}}\rangle ) \right),
\end{align*}
where $E_{\uparrow\downarrow\uparrow\uparrow}$ and $E_{\uparrow\uparrow\uparrow\downarrow}$ are the energies of their respective states  $|\uparrow\downarrow\uparrow\uparrow\rangle$ and $|\uparrow\uparrow\uparrow\downarrow\rangle$.

\vspace{0.5em}
\textbf{$\Delta E$ for system Size $L=4$ :} For a system of size $L=4$, we can express the evolved state analytically after both even and odd numbers of kicks.

After an even number of periods $n = 2m$, the state is given by
\begin{align*}
   \tilde{U}_F^{2m} |x_{\text{min}}\rangle 
   &= e^{-i T E_{\text{min}}} (-i)^{mL} \Big( e^{-i T E_{\text{max}}} |x_{\text{min}}\rangle \\
   &\quad + i e \frac{\pi}{2} \left( e^{-i T E_{\uparrow\uparrow\uparrow\uparrow}} + m e^{-i T E_{\uparrow\downarrow\uparrow\uparrow}} \right) |\uparrow\uparrow\uparrow\downarrow\rangle \\
   &\quad + i e \frac{\pi}{2} \left( e^{-i T E_{\uparrow\uparrow\uparrow\uparrow}} + m e^{-i T E_{\uparrow\uparrow\uparrow\downarrow}} \right) |\uparrow\downarrow\uparrow\uparrow\rangle \Big).
\end{align*}

After an odd number of periods $n = 2m - 1$, the state can be expressed as

\begin{align*}
   \tilde{U}_F^{2m-1}|x_{min}\rangle &= e^{-iE_{\text{min}}T} (-i)^{(2m-1)L/2} \Big(|x_{max}\rangle \\
   &+ ie\frac{\pi}{2}(m+e^{-i(E_{\uparrow\uparrow\uparrow\uparrow}+ E_{\uparrow\uparrow\uparrow\downarrow})T})|\uparrow\downarrow\uparrow\uparrow\rangle\\
   &+ ie\frac{\pi}{2}(m+e^{-i(E_{\uparrow\uparrow\uparrow\uparrow}+ E_{\uparrow\downarrow\uparrow\uparrow})T})|\uparrow\uparrow\uparrow\downarrow\rangle\Big),
\end{align*}

Finally, the energy stored in the system after an odd number of cycles is
\begin{align*}
  E^{2m-1} &= Tr[\rho H_B] - Tr[\rho_0 H_B] = E_{\text{max}} - E_{\text{min}} + \nonumber \\
  &e^2 \frac{\pi^2}{4} [1+m^2+2m\text{cos}((E_{\uparrow\uparrow\uparrow\uparrow}+ E_{\uparrow\uparrow\uparrow\downarrow})T)] E_{\uparrow\downarrow\uparrow\uparrow}\\
  &+e^2 \frac{\pi^2}{4} [1+m^2+2m\text{cos}((E_{\uparrow\uparrow\uparrow\uparrow}+ E_{\uparrow\downarrow\uparrow\uparrow})T)] E_{\uparrow\uparrow\uparrow\downarrow}.
\end{align*}

{\color{black}{\emph{Imperfection Effect of $\Delta E$.} In the main text, we discussed how the DTC can be utilized as a quantum battery by considering an imperfection parameter $e = 0.01$. However, it is important to examine how the stored energy $\Delta E$ depends on the imperfection strength $e$. This dependence allows us to evaluate the robustness and stability of the quantum battery performance against deviations from ideal conditions. In Fig.~\ref{fig:battery_different_e}, we present the plots of $\Delta E$ versus $\omega$ for different values of $e$, specifically $e \in \{0.02, 0.03, 0.04, 0.05, 0.08\}$.

The plots show that, for different values of $e$, $\Delta E$ exhibits qualitatively similar behavior. Initially, $\Delta E$ remains in a plateau region corresponding to the DTC phase. Beyond a particular value of $\omega = \omega_c$ (see main text), $\Delta E$ begins to oscillate, indicating the transition to the non-DTC phase. Notably, our results indicate that the transition point $\omega_c$ from the DTC to the non-DTC phase does not change with varying $e$.

{\color{black}{ In Fig.~\ref{fig:battery_different_alpha}, we present the stored energy $\Delta E$ (which, in our case, is equivalent to the ergotropy $\mathcal{E}$) as a function of $\omega$ for different values of $\alpha \in \{0.5,\,1.0,\,1.5\}$ with system size $L=10$ and pulse imperfection $e=0.01$. The results show that the stored energy, or equivalently the ergotropy, increases with increasing $\alpha$. Moreover, for all values of $\alpha$, the plots exhibit qualitatively similar behaviour: the stored energy initially remains nearly stable within the DTC regime, while beyond a critical value $\omega_c$ the energy begins to oscillate, indicating the breakdown of the DTC phase and the onset of the non-DTC regime. }}

\subsection{Role of initial state and imperfection effect for DTC sensor} 

The effect of the initial state and imperfections on the DTC sensor is illustrated in Fig.~\ref{fig:sensing_diff_initial_st}(a) and (b), where we present $F_Q$ as a function of $\omega$ for $\alpha = 0.5$. In  Fig.~\ref{fig:sensing_diff_initial_st}(a), we consider a fixed initial state $|\uparrow\downarrow\uparrow\downarrow\dots\rangle$, while in Fig.~\ref{fig:sensing_diff_initial_st}(b) a randomly chosen product state is considered. In both cases, results are shown for different imperfection strengths $e \in \{0.02, 0.05, 0.08, 0.1\}$. The plots reveal that as $e$ increases, the value of $F_Q$ increases; however, the qualitative behavior of the QFI remains unchanged. In particular, the transition from the DTC phase (flat region) to the non-DTC phase (oscillatory region) is independent of $e$, and the transition point $\omega_c$ does not shift with increasing imperfection.

For $\alpha = 1.5$, a similar analysis is presented in Fig.~\ref{fig:sensing_diff_initial_st}(c) and (d). In Fig.~\ref{fig:sensing_diff_initial_st}(c), the initial state is taken as $|\uparrow\downarrow\uparrow\downarrow\dots\rangle$, whereas in Fig.~\ref{fig:sensing_diff_initial_st}(d), a random product state is considered, using the same values of $e$ as before. A similar trend is observed: $F_Q$ increases with increasing $e$, while its qualitative features remain unaffected. For a larger system size $L=20$, we numerically calculate $F_Q$ as a function of $\omega$ using DMRG. The results are shown in Fig.~\ref{fig:sensing_diff_e}(a) and (b) for $\alpha = 0.5$ and $\alpha = 1.5$, respectively. The plots again demonstrate that $F_Q$ increases with increasing $e$ for both values of $\alpha$.

}}

\bibliography{References}

@article{Wilczek2012,
  title = {Quantum Time Crystals},
  author = {Wilczek, Frank},
  journal = {Phys. Rev. Lett.},
  volume = {109},
  issue = {16},
  pages = {160401},
  numpages = {5},
  year = {2012},
  month = {Oct},
  publisher = {American Physical Society},
  doi = {10.1103/PhysRevLett.109.160401},
  url = {https://link.aps.org/doi/10.1103/PhysRevLett.109.160401}
}

@article{Li2012,
  title = {Space-Time Crystals of Trapped Ions},
  author = {Li, Tongcang and Gong, Zhe-Xuan and Yin, Zhang-Qi and Quan, H. T. and Yin, Xiaobo and Zhang, Peng and Duan, L.-M. and Zhang, Xiang},
  journal = {Phys. Rev. Lett.},
  volume = {109},
  issue = {16},
  pages = {163001},
  numpages = {5},
  year = {2012},
  month = {Oct},
  publisher = {American Physical Society},
  doi = {10.1103/PhysRevLett.109.163001},
  url = {https://link.aps.org/doi/10.1103/PhysRevLett.109.163001}
}

@article{Wilczek2013,
  title = {Superfluidity and Space-Time Translation Symmetry Breaking},
  author = {Wilczek, Frank},
  journal = {Phys. Rev. Lett.},
  volume = {111},
  issue = {25},
  pages = {250402},
  numpages = {4},
  year = {2013},
  month = {Dec},
  publisher = {American Physical Society},
  doi = {10.1103/PhysRevLett.111.250402},
  url = {https://link.aps.org/doi/10.1103/PhysRevLett.111.250402}
}

@article{Alicki2013,
  title = {Entanglement boost for extractable work from ensembles of quantum batteries},
  author = {Alicki, Robert and Fannes, Mark},
  journal = {Phys. Rev. E},
  volume = {87},
  issue = {4},
  pages = {042123},
  numpages = {4},
  year = {2013},
  month = {Apr},
  publisher = {American Physical Society},
  doi = {10.1103/PhysRevE.87.042123},
  url = {https://link.aps.org/doi/10.1103/PhysRevE.87.042123}
}

@article{Rossini2019b,
  title = {Many-body localized quantum batteries},
  author = {Rossini, Davide and Andolina, Gian Marcello and Polini, Marco},
  journal = {Phys. Rev. B},
  volume = {100},
  issue = {11},
  pages = {115142},
  numpages = {11},
  year = {2019},
  month = {Sep},
  publisher = {American Physical Society},
  doi = {10.1103/PhysRevB.100.115142},
  url = {https://link.aps.org/doi/10.1103/PhysRevB.100.115142}
}

@article{Choi_Nature_17,
  title = {Observation of discrete time-crystalline order in a disordered dipolar many-body system},
  author = {Soonwon Choi and Joonhee Choi and Renate Landig and Georg Kucsko and Hengyun Zhou and  Junichi Isoya and Fedor Jelezko and Shinobu Onoda and Hitoshi Sumiya and Vedika Khemani and Curt von Keyserlingk and Norman Y. Yao and Eugene Demlerm and Mikhail D. Lukin},
  journal = {Nature},
  volume = {543},
  pages = {221},
  year = {2017},
 }

@article{hess_Nature_17,
  title = {Observation of a discrete time crystal},
  author = {J. Zhang and P. W. Hess and A. Kyprianidis and P. Becker and A. Lee and J. Smith and G. Pagano and I.-D. Potirniche and A. C. Potter and A. Vishwanath and N. Y. Yao  and C. Monroe },
  journal = {Nature},
  volume = {543},
 pages = {217},
  year = {2017},
 }

@article{Zhan_Nature_22,
  title = {Digital quantum simulation of Floquet symmetry-protected topological phases},
  author = {Xu Zhang and Wenjie Jiang and Jinfeng Deng and Ke Wang and Jiachen Chen and Pengfei Zhang and Wenhui Ren and Hang Dong and Shibo Xu and Yu Gao and Feitong Jin and Xuhao Zhu and Qiujiang Guo and Hekang Li and Chao Song and Alexey V. Gorshkov and Thomas Iadecola and Fangli Liu and Zhe-Xuan Gong and Zhen Wang and Dong-Ling Deng  and H. Wang  },
  journal = {Nature},
  volume = {607},
  pages = {468},
  year = {2022},
 }

@article{science_adv_20,
  title = {Simulating complex quantum networks with time crystals},
  author = {M. P. Estarellas1 and T. Osada1 and V. M. Bastidas and B. Renoust and K. Sanaka and W. J. Munro1 and K. Nemoto},
  journal = {Science Advances},
  volume = {6},
  pages = {eaay8892},
  year = {2020},
 }

@article{Arjmandi2023,
  title = {Localization effects in disordered quantum batteries},
  author = {Arjmandi, Mohammad B. and Mohammadi, Hamidreza and Saguia, Andreia and Sarandy, Marcelo S. and Santos, Alan C.},
  journal = {Phys. Rev. E},
  volume = {108},
  issue = {6},
  pages = {064106},
  numpages = {9},
  year = {2023},
  month = {Dec},
  publisher = {American Physical Society},
  doi = {10.1103/PhysRevE.108.064106},
  url = {https://link.aps.org/doi/10.1103/PhysRevE.108.064106}
}

@article{Bomantara2018,
  title = {Simulation of Non-Abelian Braiding in Majorana Time Crystals},
  author = {Bomantara, Raditya Weda and Gong, Jiangbin},
  journal = {Phys. Rev. Lett.},
  volume = {120},
  issue = {23},
  pages = {230405},
  numpages = {5},
  year = {2018},
  month = {Jun},
  publisher = {American Physical Society},
  doi = {10.1103/PhysRevLett.120.230405},
  url = {https://link.aps.org/doi/10.1103/PhysRevLett.120.230405}
}

@article{PhysRevLett.125.240602,
  title = {Nonequilibrium Many-Body Quantum Engine Driven by Time-Translation Symmetry Breaking},
  author = {Carollo, Federico and Brandner, Kay and Lesanovsky, Igor},
  journal = {Phys. Rev. Lett.},
  volume = {125},
  issue = {24},
  pages = {240602},
  numpages = {6},
  year = {2020},
  month = {Dec},
  publisher = {American Physical Society},
  doi = {10.1103/PhysRevLett.125.240602},
  url = {https://link.aps.org/doi/10.1103/PhysRevLett.125.240602}
}

@article{Zhu2023,
  title = {Charging Quantum Batteries via Indefinite Causal Order: Theory and Experiment},
  author = {Zhu, Gaoyan and Chen, Yuanbo and Hasegawa, Yoshihiko and Xue, Peng},
  journal = {Phys. Rev. Lett.},
  volume = {131},
  issue = {24},
  pages = {240401},
  numpages = {7},
  year = {2023},
  month = {Dec},
  publisher = {American Physical Society},
  doi = {10.1103/PhysRevLett.131.240401},
  url = {https://link.aps.org/doi/10.1103/PhysRevLett.131.240401}
}

@article{Iemini2024,
  title = {Floquet time crystals as quantum sensors of ac fields},
  author = {Iemini, Fernando and Fazio, Rosario and Sanpera, Anna},
  journal = {Phys. Rev. A},
  volume = {109},
  issue = {5},
  pages = {L050203},
  numpages = {6},
  year = {2024},
  month = {May},
  publisher = {American Physical Society},
  doi = {10.1103/PhysRevA.109.L050203},
  url = {https://link.aps.org/doi/10.1103/PhysRevA.109.L050203}
}

@article{Cabot2024,
  title = {Continuous Sensing and Parameter Estimation with the Boundary Time Crystal},
  author = {Cabot, Albert and Carollo, Federico and Lesanovsky, Igor},
  journal = {Phys. Rev. Lett.},
  volume = {132},
  issue = {5},
  pages = {050801},
  numpages = {6},
  year = {2024},
  month = {Jan},
  publisher = {American Physical Society},
  doi = {10.1103/PhysRevLett.132.050801},
  url = {https://link.aps.org/doi/10.1103/PhysRevLett.132.050801}
}

@article{Hogan_2024,
doi = {10.1088/1402-4896/ad2efd},
url = {https://dx.doi.org/10.1088/1402-4896/ad2efd},
year = {2024},
month = {apr},
publisher = {IOP Publishing},
volume = {99},
number = {5},
pages = {055118},
author = {Hogan, Andrew R and Martin, Andy M},
title = {Quench dynamics in the Jaynes-Cummings-Hubbard and Dicke models},
journal = {Physica Scripta}
}

@article{Mondal2022,
  title = {Periodically driven many-body quantum battery},
  author = {Mondal, Saikat and Bhattacharjee, Sourav},
  journal = {Phys. Rev. E},
  volume = {105},
  issue = {4},
  pages = {044125},
  numpages = {8},
  year = {2022},
  month = {Apr},
  publisher = {American Physical Society},
  doi = {10.1103/PhysRevE.105.044125},
  url = {https://link.aps.org/doi/10.1103/PhysRevE.105.044125}
}

@article{Bukov04032015,
author = {Marin Bukov, Luca D'Alessio and Anatoli Polkovnikov},
title = {Universal high-frequency behavior of periodically driven systems: from dynamical stabilization to Floquet engineering},
journal = {Advances in Physics},
volume = {64},
number = {2},
pages = {139--226},
year = {2015},
publisher = {Taylor \& Francis},
doi = {10.1080/00018732.2015.1055918},


URL = { 
    
        https://doi.org/10.1080/00018732.2015.1055918
    
    

},
eprint = { 
    
        https://doi.org/10.1080/00018732.2015.1055918
    
    

}

}

@article{Eckardt2017,
  title = {Colloquium: Atomic quantum gases in periodically driven optical lattices},
  author = {Eckardt, Andr\'e},
  journal = {Rev. Mod. Phys.},
  volume = {89},
  issue = {1},
  pages = {011004},
  numpages = {30},
  year = {2017},
  month = {Mar},
  publisher = {American Physical Society},
  doi = {10.1103/RevModPhys.89.011004},
  url = {https://link.aps.org/doi/10.1103/RevModPhys.89.011004}
}

@article{Ghosh2022,
  title = {Dimensional enhancements in a quantum battery with imperfections},
  author = {Ghosh, Srijon and Sen(De), Aditi},
  journal = {Phys. Rev. A},
  volume = {105},
  issue = {2},
  pages = {022628},
  numpages = {10},
  year = {2022},
  month = {Feb},
  publisher = {American Physical Society},
  doi = {10.1103/PhysRevA.105.022628},
  url = {https://link.aps.org/doi/10.1103/PhysRevA.105.022628}
}

@article{Juli2020,
  title = {Bounds on the capacity and power of quantum batteries},
  author = {Juli\`a-Farr\'e, Sergi and Salamon, Tymoteusz and Riera, Arnau and Bera, Manabendra N. and Lewenstein, Maciej},
  journal = {Phys. Rev. Res.},
  volume = {2},
  issue = {2},
  pages = {023113},
  numpages = {16},
  year = {2020},
  month = {May},
  publisher = {American Physical Society},
  doi = {10.1103/PhysRevResearch.2.023113},
  url = {https://link.aps.org/doi/10.1103/PhysRevResearch.2.023113}
}

@article{Ferraro2018,
  title = {High-Power Collective Charging of a Solid-State Quantum Battery},
  author = {Ferraro, Dario and Campisi, Michele and Andolina, Gian Marcello and Pellegrini, Vittorio and Polini, Marco},
  journal = {Phys. Rev. Lett.},
  volume = {120},
  issue = {11},
  pages = {117702},
  numpages = {6},
  year = {2018},
  month = {Mar},
  publisher = {American Physical Society},
  doi = {10.1103/PhysRevLett.120.117702},
  url = {https://link.aps.org/doi/10.1103/PhysRevLett.120.117702}
}

@article{Andolina2019,
  title = {Quantum versus classical many-body batteries},
  author = {Andolina, Gian Marcello and Keck, Maximilian and Mari, Andrea and Giovannetti, Vittorio and Polini, Marco},
  journal = {Phys. Rev. B},
  volume = {99},
  issue = {20},
  pages = {205437},
  numpages = {7},
  year = {2019},
  month = {May},
  publisher = {American Physical Society},
  doi = {10.1103/PhysRevB.99.205437},
  url = {https://link.aps.org/doi/10.1103/PhysRevB.99.205437}
}

@article{Santos2019,
  title = {Stable adiabatic quantum batteries},
  author = {Santos, Alan C. and \ifmmode \mbox{\c{C}}\else \c{C}\fi{}akmak, Bar\ifmmode \imath \else \i \fi{}\ifmmode \mbox{\c{s}}\else \c{s}\fi{} and Campbell, Steve and Zinner, Nikolaj T.},
  journal = {Phys. Rev. E},
  volume = {100},
  issue = {3},
  pages = {032107},
  numpages = {8},
  year = {2019},
  month = {Sep},
  publisher = {American Physical Society},
  doi = {10.1103/PhysRevE.100.032107},
  url = {https://link.aps.org/doi/10.1103/PhysRevE.100.032107}
}

@article{Rossini2020,
  title = {Quantum Advantage in the Charging Process of Sachdev-Ye-Kitaev Batteries},
  author = {Rossini, Davide and Andolina, Gian Marcello and Rosa, Dario and Carrega, Matteo and Polini, Marco},
  journal = {Phys. Rev. Lett.},
  volume = {125},
  issue = {23},
  pages = {236402},
  numpages = {6},
  year = {2020},
  month = {Dec},
  publisher = {American Physical Society},
  doi = {10.1103/PhysRevLett.125.236402},
  url = {https://link.aps.org/doi/10.1103/PhysRevLett.125.236402}
}

@article{Sen2021,
  title = {Local passivity and entanglement in shared quantum batteries},
  author = {Sen, Kornikar and Sen, Ujjwal},
  journal = {Phys. Rev. A},
  volume = {104},
  issue = {3},
  pages = {L030402},
  numpages = {6},
  year = {2021},
  month = {Sep},
  publisher = {American Physical Society},
  doi = {10.1103/PhysRevA.104.L030402},
  url = {https://link.aps.org/doi/10.1103/PhysRevA.104.L030402}
}

@article{Crescente2020,
  title = {Ultrafast charging in a two-photon Dicke quantum battery},
  author = {Crescente, Alba and Carrega, Matteo and Sassetti, Maura and Ferraro, Dario},
  journal = {Phys. Rev. B},
  volume = {102},
  issue = {24},
  pages = {245407},
  numpages = {12},
  year = {2020},
  month = {Dec},
  publisher = {American Physical Society},
  doi = {10.1103/PhysRevB.102.245407},
  url = {https://link.aps.org/doi/10.1103/PhysRevB.102.245407}
}

@article{Crescente2022,
  title = {Enhancing coherent energy transfer between quantum devices via a mediator},
  author = {Crescente, Alba and Ferraro, Dario and Carrega, Matteo and Sassetti, Maura},
  journal = {Phys. Rev. Res.},
  volume = {4},
  issue = {3},
  pages = {033216},
  numpages = {16},
  year = {2022},
  month = {Sep},
  publisher = {American Physical Society},
  doi = {10.1103/PhysRevResearch.4.033216},
  url = {https://link.aps.org/doi/10.1103/PhysRevResearch.4.033216}
}

@article{puri2024floquet,
  title={Floquet driven long-range interactions induce super-extensive scaling in quantum battery},
  author={Puri, Stavya and Konar, Tanoy Kanti and Lakkaraju, Leela Ganesh Chandra and De, Aditi Sen},
  journal={arXiv preprint arXiv:2412.00921},
  year={2024}
}

@article{Appel2009,
  title = {Mesoscopic atomic entanglement for precision measurements beyond the standard quantum limit},
  volume = {106},
  ISSN = {1091-6490},
  url = {http://dx.doi.org/10.1073/pnas.0901550106},
  DOI = {10.1073/pnas.0901550106},
  number = {27},
  journal = {Proceedings of the National Academy of Sciences},
  publisher = {Proceedings of the National Academy of Sciences},
  author = {Appel,  J. and Windpassinger,  P. J. and Oblak,  D. and Hoff,  U. B. and Kjærgaard,  N. and Polzik,  E. S.},
  year = {2009},
  month = jul,
  pages = {10960–10965}
}

@article{LouchetChauvet2010,
  title = {Entanglement-assisted atomic clock beyond the projection noise limit},
  volume = {12},
  ISSN = {1367-2630},
  url = {http://dx.doi.org/10.1088/1367-2630/12/6/065032},
  DOI = {10.1088/1367-2630/12/6/065032},
  number = {6},
  journal = {New Journal of Physics},
  publisher = {IOP Publishing},
  author = {Louchet-Chauvet,  Anne and Appel,  J\"{u}rgen and Renema,  Jelmer J and Oblak,  Daniel and Kjaergaard,  Niels and Polzik,  Eugene S},
  year = {2010},
  month = jun,
  pages = {065032}
}

@article{Mitchell2004,
  title = {Super-resolving phase measurements with a multiphoton entangled state},
  volume = {429},
  ISSN = {1476-4687},
  url = {http://dx.doi.org/10.1038/nature02493},
  DOI = {10.1038/nature02493},
  number = {6988},
  journal = {Nature},
  publisher = {Springer Science and Business Media LLC},
  author = {Mitchell,  M. W. and Lundeen,  J. S. and Steinberg,  A. M.},
  year = {2004},
  month = may,
  pages = {161–164}
}

@article{Nagata2007,
  title = {Beating the Standard Quantum Limit with Four-Entangled Photons},
  volume = {316},
  ISSN = {1095-9203},
  url = {http://dx.doi.org/10.1126/science.1138007},
  DOI = {10.1126/science.1138007},
  number = {5825},
  journal = {Science},
  publisher = {American Association for the Advancement of Science (AAAS)},
  author = {Nagata,  Tomohisa and Okamoto,  Ryo and O’Brien,  Jeremy L. and Sasaki,  Keiji and Takeuchi,  Shigeki},
  year = {2007},
  month = may,
  pages = {726–729}
}

@article{Wasilewski2010,
  title = {Quantum Noise Limited and Entanglement-Assisted Magnetometry},
  author = {Wasilewski, W. and Jensen, K. and Krauter, H. and Renema, J. J. and Balabas, M. V. and Polzik, E. S.},
  journal = {Phys. Rev. Lett.},
  volume = {104},
  issue = {13},
  pages = {133601},
  numpages = {4},
  year = {2010},
  month = {Mar},
  publisher = {American Physical Society},
  doi = {10.1103/PhysRevLett.104.133601},
  url = {https://link.aps.org/doi/10.1103/PhysRevLett.104.133601}
}

@article{Sewell2012,
  title = {Magnetic Sensitivity Beyond the Projection Noise Limit by Spin Squeezing},
  author = {Sewell, R. J. and Koschorreck, M. and Napolitano, M. and Dubost, B. and Behbood, N. and Mitchell, M. W.},
  journal = {Phys. Rev. Lett.},
  volume = {109},
  issue = {25},
  pages = {253605},
  numpages = {5},
  year = {2012},
  month = {Dec},
  publisher = {American Physical Society},
  doi = {10.1103/PhysRevLett.109.253605},
  url = {https://link.aps.org/doi/10.1103/PhysRevLett.109.253605}
}

@article{Roos2006,
  title = {‘Designer atoms’ for quantum metrology},
  volume = {443},
  ISSN = {1476-4687},
  url = {http://dx.doi.org/10.1038/nature05101},
  DOI = {10.1038/nature05101},
  number = {7109},
  journal = {Nature},
  publisher = {Springer Science and Business Media LLC},
  author = {Roos,  C. F. and Chwalla,  M. and Kim,  K. and Riebe,  M. and Blatt,  R.},
  year = {2006},
  month = sep,
  pages = {316–319}
}

@article{Leibfried2004,
  title = {rrd Heisenberg-Limited Spectroscopy with Multiparticle Entangled States},
  volume = {304},
  ISSN = {1095-9203},
  url = {http://dx.doi.org/10.1126/science.1097576},
  
  number = {5676},
  journal = {Science},
  publisher = {American Association for the Advancement of Science (AAAS)},
  author = {Leibfried,  D. and Barrett,  M. D. and Schaetz,  T. and Britton,  J. and Chiaverini,  J. and Itano,  W. M. and Jost,  J. D. and Langer,  C. and Wineland,  D. J.},
  year = {2004},
  month = jun,
  pages = {1476–1478}
}

@article{Mishra2021a,
  title = {Driving Enhanced Quantum Sensing in Partially Accessible Many-Body Systems},
  author = {Mishra, Utkarsh and Bayat, Abolfazl},
  journal = {Phys. Rev. Lett.},
  volume = {127},
  issue = {8},
  pages = {080504},
  numpages = {7},
  year = {2021},
  month = {Aug},
  publisher = {American Physical Society},
  doi = {10.1103/PhysRevLett.127.080504},
  url = {https://link.aps.org/doi/10.1103/PhysRevLett.127.080504}
}

@article{agarwal2025,
  title={Critical quantum metrology using non-Hermitian spin model with RT-symmetry},
  author={Agarwal, Keshav Das and Konar, Tanoy Kanti and Lakkaraju, Leela Ganesh Chandra and De, Aditi Sen},
  journal={arXiv:2503.24331},
  url={https://arxiv.org/abs/2503.24331},
  year={2025}
}

@article{Yousefjani2025,
  title = {Discrete time crystal phase as a resource for quantum-enhanced sensing},
  author = {Yousefjani, Rozhin and Sacha, Krzysztof and Bayat, Abolfazl},
  journal = {Phys. Rev. B},
  volume = {111},
  issue = {12},
  pages = {125159},
  numpages = {14},
  year = {2025},
  month = {Mar},
  publisher = {American Physical Society},
  doi = {10.1103/PhysRevB.111.125159},
  url = {https://link.aps.org/doi/10.1103/PhysRevB.111.125159}
}

@article{biswas2025,
  title={Discrete Time Crystals in the spin-s Central Spin Model},
  author={Biswas, Hillol and Choudhury, Sayan},
  journal={arXiv:2505.13207},
  url={https://arxiv.org/abs/2505.13207},
  year={2025}
}

@article{biswas2025floquet,
  title={The Floquet central spin model: A platform to realize eternal time crystals, entanglement steering, and multiparameter metrology},
  author={Biswas, Hillol and Choudhury, Sayan},
  journal={arXiv:2501.18472},
  url={https://arxiv.org/abs/2501.18472},
  year={2025}
}

@article{Sahoo2024a,
  title = {Localization-driven quantum sensing},
  author = {Sahoo, Ayan and Mishra, Utkarsh and Rakshit, Debraj},
  journal = {Phys. Rev. A},
  volume = {109},
  issue = {3},
  pages = {L030601},
  numpages = {8},
  year = {2024},
  month = {Mar},
  publisher = {American Physical Society},
  doi = {10.1103/PhysRevA.109.L030601},
  url = {https://link.aps.org/doi/10.1103/PhysRevA.109.L030601}
}

@article{Sahoo2025a,
  title = {Stark localization near Aubry-Andr\'e criticality},
  author = {Sahoo, Ayan and Saha, Aitijhya and Rakshit, Debraj},
  journal = {Phys. Rev. B},
  volume = {111},
  issue = {2},
  pages = {024205},
  numpages = {12},
  year = {2025},
  month = {Jan},
  publisher = {American Physical Society},
  doi = {10.1103/PhysRevB.111.024205},
  url = {https://link.aps.org/doi/10.1103/PhysRevB.111.024205}
}

@article{mondal2024,
  title = {Multicritical quantum sensors driven by symmetry breaking},
  author = {Mondal, Sayan and Sahoo, Ayan and Sen, Ujjwal and Rakshit, Debraj},
  journal = {Phys. Rev. B},
  volume = {112},
  issue = {23},
  pages = {235165},
  numpages = {13},
  year = {2025},
  month = {Dec},
  publisher = {American Physical Society},
  doi = {10.1103/18jh-9zc4},
  url = {https://link.aps.org/doi/10.1103/18jh-9zc4}
}

@article{sahoo2024b,
  title = {Enhanced sensing of a weak Stark field under the influence of Aubry-Andr\'e-Harper criticality},
  author = {Sahoo, Ayan and Rakshit, Debraj},
  journal = {Phys. Rev. A},
  volume = {113},
  issue = {2},
  pages = {022601},
  numpages = {11},
  year = {2026},
  month = {Feb},
  publisher = {American Physical Society},
  doi = {10.1103/kq1w-pgb8},
  url = {https://link.aps.org/doi/10.1103/kq1w-pgb8}
}

@article{He2023,
  title = {Stark Localization as a Resource for Weak-Field Sensing with Super-Heisenberg Precision},
  author = {He, Xingjian and Yousefjani, Rozhin and Bayat, Abolfazl},
  journal = {Phys. Rev. Lett.},
  volume = {131},
  issue = {1},
  pages = {010801},
  numpages = {7},
  year = {2023},
  month = {Jul},
  publisher = {American Physical Society},
  doi = {10.1103/PhysRevLett.131.010801},
  url = {https://link.aps.org/doi/10.1103/PhysRevLett.131.010801}
}

@article{debnath2025tilt,
  title={Tilt-Induced Localization in Interacting Bose-Einstein Condensates for Quantum Sensing},
  author={Debnath, Argha and Gajda, Mariusz and Rakshit, Debraj},
  journal={arXiv:2506.06173},
  url={https://arxiv.org/abs/2506.06173},
  year={2025}
}

@article{Montenegro2023,
  title = {Quantum metrology with boundary time crystals},
  volume = {6},
  ISSN = {2399-3650},
  url = {http://dx.doi.org/10.1038/s42005-023-01423-6},
  
  number = {1},
  journal = {Communications Physics},
  publisher = {Springer Science and Business Media LLC},
  author = {Montenegro,  Victor and Genoni,  Marco G. and Bayat,  Abolfazl and Paris,  Matteo G. A.},
  year = {2023},
  month = oct 
}

@article{Montenegro2025,
  title = {Review: Quantum metrology and sensing with many-body systems},
  volume = {1134},
  ISSN = {0370-1573},
  url = {http://dx.doi.org/10.1016/j.physrep.2025.05.005},
  DOI = {10.1016/j.physrep.2025.05.005},
  journal = {Physics Reports},
  publisher = {Elsevier BV},
  author = {Montenegro,  Victor and Mukhopadhyay,  Chiranjib and Yousefjani,  Rozhin and Sarkar,  Saubhik and Mishra,  Utkarsh and Paris,  Matteo G.A. and Bayat,  Abolfazl},
  year = {2025},
  month = jul,
  pages = {1–62}
}

@article{Rams2018,
  title = {At the Limits of Criticality-Based Quantum Metrology: Apparent Super-Heisenberg Scaling Revisited},
  author = {Rams, Marek M. and Sierant, Piotr and Dutta, Omyoti and Horodecki, Pawe\l{} and Zakrzewski, Jakub},
  journal = {Phys. Rev. X},
  volume = {8},
  issue = {2},
  pages = {021022},
  numpages = {16},
  year = {2018},
  month = {Apr},
  publisher = {American Physical Society},
  doi = {10.1103/PhysRevX.8.021022},
  url = {https://link.aps.org/doi/10.1103/PhysRevX.8.021022}
}

@article{Mishra2022,
  title = {Integrable quantum many-body sensors for AC field sensing},
  volume = {12},
  ISSN = {2045-2322},
  url = {http://dx.doi.org/10.1038/s41598-022-17381-y},
  
  number = {1},
  journal = {Scientific Reports},
  publisher = {Springer Science and Business Media LLC},
  author = {Mishra,  Utkarsh and Bayat,  Abolfazl},
  year = {2022},
  month = aug 
}

@article{Dooley2021,
  title = {Robust Quantum Sensing in Strongly Interacting Systems with Many-Body Scars},
  author = {Dooley, Shane},
  journal = {PRX Quantum},
  volume = {2},
  issue = {2},
  pages = {020330},
  numpages = {12},
  year = {2021},
  month = {May},
  publisher = {American Physical Society},
  doi = {10.1103/PRXQuantum.2.020330},
  url = {https://link.aps.org/doi/10.1103/PRXQuantum.2.020330}
}

@article{Desaules2022,
  title = {Extensive Multipartite Entanglement from su(2) Quantum Many-Body Scars},
  author = {Desaules, Jean-Yves and Pietracaprina, Francesca and Papi\ifmmode \acute{c}\else \'{c}\fi{}, Zlatko and Goold, John and Pappalardi, Silvia},
  journal = {Phys. Rev. Lett.},
  volume = {129},
  issue = {2},
  pages = {020601},
  numpages = {7},
  year = {2022},
  month = {Jul},
  publisher = {American Physical Society},
  doi = {10.1103/PhysRevLett.129.020601},
  url = {https://link.aps.org/doi/10.1103/PhysRevLett.129.020601}
}

@article{Dooley2023b,
  title = {Entanglement enhanced metrology with quantum many-body scars},
  author = {Dooley, Shane and Pappalardi, Silvia and Goold, John},
  journal = {Phys. Rev. B},
  volume = {107},
  issue = {3},
  pages = {035123},
  numpages = {7},
  year = {2023},
  month = {Jan},
  publisher = {American Physical Society},
  doi = {10.1103/PhysRevB.107.035123},
  url = {https://link.aps.org/doi/10.1103/PhysRevB.107.035123}
}

@article{Guo2023,
  title = {Origin of Hilbert-space quantum scars in unconstrained models},
  author = {Guo, Zexian and Liu, Bobo and Gao, Yu and Yang, Ang and Wang, Junlin and Ma, Jinlou and Ying, Lei},
  journal = {Phys. Rev. B},
  volume = {108},
  issue = {7},
  pages = {075124},
  numpages = {10},
  year = {2023},
  month = {Aug},
  publisher = {American Physical Society},
  doi = {10.1103/PhysRevB.108.075124},
  url = {https://link.aps.org/doi/10.1103/PhysRevB.108.075124}
}

@article{Liu2023,
  title = {Discrete Time Crystal Enabled by Stark Many-Body Localization},
  author = {Liu, Shuo and Zhang, Shi-Xin and Hsieh, Chang-Yu and Zhang, Shengyu and Yao, Hong},
  journal = {Phys. Rev. Lett.},
  volume = {130},
  issue = {12},
  pages = {120403},
  numpages = {8},
  year = {2023},
  month = {Mar},
  publisher = {American Physical Society},
  doi = {10.1103/PhysRevLett.130.120403},
  url = {https://link.aps.org/doi/10.1103/PhysRevLett.130.120403}
}

@article{Zanardi2008,
  title = {Quantum criticality as a resource for quantum estimation},
  author = {Zanardi, Paolo and Paris, Matteo G. A. and Campos Venuti, Lorenzo},
  journal = {Phys. Rev. A},
  volume = {78},
  issue = {4},
  pages = {042105},
  numpages = {7},
  year = {2008},
  month = {Oct},
  publisher = {American Physical Society},
  doi = {10.1103/PhysRevA.78.042105},
  url = {https://link.aps.org/doi/10.1103/PhysRevA.78.042105}
}

@article{Campos2007,
  title = {Quantum Critical Scaling of the Geometric Tensors},
  author = {Campos Venuti, Lorenzo and Zanardi, Paolo},
  journal = {Phys. Rev. Lett.},
  volume = {99},
  issue = {9},
  pages = {095701},
  numpages = {4},
  year = {2007},
  month = {Aug},
  publisher = {American Physical Society},
  doi = {10.1103/PhysRevLett.99.095701},
  url = {https://link.aps.org/doi/10.1103/PhysRevLett.99.095701}
}

@article{Tsang2013,
  title = {Quantum transition-edge detectors},
  author = {Tsang, Mankei},
  journal = {Phys. Rev. A},
  volume = {88},
  issue = {2},
  pages = {021801},
  numpages = {5},
  year = {2013},
  month = {Aug},
  publisher = {American Physical Society},
  doi = {10.1103/PhysRevA.88.021801},
  url = {https://link.aps.org/doi/10.1103/PhysRevA.88.021801}
}

@article{Garbe2020,
  title = {Critical Quantum Metrology with a Finite-Component Quantum Phase Transition},
  author = {Garbe, Louis and Bina, Matteo and Keller, Arne and Paris, Matteo G. A. and Felicetti, Simone},
  journal = {Phys. Rev. Lett.},
  volume = {124},
  issue = {12},
  pages = {120504},
  numpages = {5},
  year = {2020},
  month = {Mar},
  publisher = {American Physical Society},
  doi = {10.1103/PhysRevLett.124.120504},
  url = {https://link.aps.org/doi/10.1103/PhysRevLett.124.120504}
}

@article{Chu2021,
  title = {Dynamic Framework for Criticality-Enhanced Quantum Sensing},
  author = {Chu, Yaoming and Zhang, Shaoliang and Yu, Baiyi and Cai, Jianming},
  journal = {Phys. Rev. Lett.},
  volume = {126},
  issue = {1},
  pages = {010502},
  numpages = {7},
  year = {2021},
  month = {Jan},
  publisher = {American Physical Society},
  doi = {10.1103/PhysRevLett.126.010502},
  url = {https://link.aps.org/doi/10.1103/PhysRevLett.126.010502}
}

@article{Montenegro2021b,
  title = {Global Sensing and Its Impact for Quantum Many-Body Probes with Criticality},
  author = {Montenegro, Victor and Mishra, Utkarsh and Bayat, Abolfazl},
  journal = {Phys. Rev. Lett.},
  volume = {126},
  issue = {20},
  pages = {200501},
  numpages = {6},
  year = {2021},
  month = {May},
  publisher = {American Physical Society},
  doi = {10.1103/PhysRevLett.126.200501},
  url = {https://link.aps.org/doi/10.1103/PhysRevLett.126.200501}
}

@article{Mirkhalaf2020,
  title = {Supersensitive quantum sensor based on criticality in an antiferromagnetic spinor condensate},
  author = {Mirkhalaf, Safoura S. and Witkowska, Emilia and Lepori, Luca},
  journal = {Phys. Rev. A},
  volume = {101},
  issue = {4},
  pages = {043609},
  numpages = {13},
  year = {2020},
  month = {Apr},
  publisher = {American Physical Society},
  doi = {10.1103/PhysRevA.101.043609},
  url = {https://link.aps.org/doi/10.1103/PhysRevA.101.043609}
}

@article{Frerot2018,
  title = {Quantum Critical Metrology},
  author = {Fr\'erot, Ir\'en\'ee and Roscilde, Tommaso},
  journal = {Phys. Rev. Lett.},
  volume = {121},
  issue = {2},
  pages = {020402},
  numpages = {6},
  year = {2018},
  month = {Jul},
  publisher = {American Physical Society},
  doi = {10.1103/PhysRevLett.121.020402},
  url = {https://link.aps.org/doi/10.1103/PhysRevLett.121.020402}
}

@article{Zanardi2006,
  title = {Ground state overlap and quantum phase transitions},
  author = {Zanardi, Paolo and Paunkovi\ifmmode \acute{c}\else \'{c}\fi{}, Nikola},
  journal = {Phys. Rev. E},
  volume = {74},
  issue = {3},
  pages = {031123},
  numpages = {6},
  year = {2006},
  month = {Sep},
  publisher = {American Physical Society},
  doi = {10.1103/PhysRevE.74.031123},
  url = {https://link.aps.org/doi/10.1103/PhysRevE.74.031123}
}

@article{You2007,
  title = {Fidelity, dynamic structure factor, and susceptibility in critical phenomena},
  author = {You, Wen-Long and Li, Ying-Wai and Gu, Shi-Jian},
  journal = {Phys. Rev. E},
  volume = {76},
  issue = {2},
  pages = {022101},
  numpages = {4},
  year = {2007},
  month = {Aug},
  publisher = {American Physical Society},
  doi = {10.1103/PhysRevE.76.022101},
  url = {https://link.aps.org/doi/10.1103/PhysRevE.76.022101}
}

@article{Zanardi2007c,
  title = {Information-Theoretic Differential Geometry of Quantum Phase Transitions},
  author = {Zanardi, Paolo and Giorda, Paolo and Cozzini, Marco},
  journal = {Phys. Rev. Lett.},
  volume = {99},
  issue = {10},
  pages = {100603},
  numpages = {4},
  year = {2007},
  month = {Sep},
  publisher = {American Physical Society},
  doi = {10.1103/PhysRevLett.99.100603},
  url = {https://link.aps.org/doi/10.1103/PhysRevLett.99.100603}
}

@article{Ilias2022,
  title = {Criticality-Enhanced Quantum Sensing via Continuous Measurement},
  author = {Ilias, Theodoros and Yang, Dayou and Huelga, Susana F. and Plenio, Martin B.},
  journal = {PRX Quantum},
  volume = {3},
  issue = {1},
  pages = {010354},
  numpages = {21},
  year = {2022},
  month = {Mar},
  publisher = {American Physical Society},
  doi = {10.1103/PRXQuantum.3.010354},
  url = {https://link.aps.org/doi/10.1103/PRXQuantum.3.010354}
}

@article{Gietka2021,
  title = {Adiabatic critical quantum metrology cannot reach the Heisenberg limit even when shortcuts to adiabaticity are applied},
  volume = {5},
  ISSN = {2521-327X},
  url = {http://dx.doi.org/10.22331/q-2021-07-01-489},
  DOI = {10.22331/q-2021-07-01-489},
  journal = {Quantum},
  publisher = {Verein zur Forderung des Open Access Publizierens in den Quantenwissenschaften},
  author = {Gietka,  Karol and Metz,  Friederike and Keller,  Tim and Li,  Jing},
  year = {2021},
  month = jul,
  pages = {489}
}

@article{GU2010,
  title = {FIDELITY APPROACH TO QUANTUM PHASE TRANSITIONS},
  volume = {24},
  ISSN = {1793-6578},
  url = {http://dx.doi.org/10.1142/S0217979210056335},
  DOI = {10.1142/s0217979210056335},
  number = {23},
  journal = {International Journal of Modern Physics B},
  publisher = {World Scientific Pub Co Pte Ltd},
  author = {GU,  SHI-JIAN},
  year = {2010},
  month = sep,
  pages = {4371–4458}
}

@article{Khemani2016,
  title = {Phase Structure of Driven Quantum Systems},
  author = {Khemani, Vedika and Lazarides, Achilleas and Moessner, Roderich and Sondhi, S. L.},
  journal = {Phys. Rev. Lett.},
  volume = {116},
  issue = {25},
  pages = {250401},
  numpages = {6},
  year = {2016},
  month = {Jun},
  publisher = {American Physical Society},
  doi = {10.1103/PhysRevLett.116.250401},
  url = {https://link.aps.org/doi/10.1103/PhysRevLett.116.250401}
}

@article{Else2016,
  title = {Floquet Time Crystals},
  author = {Else, Dominic V. and Bauer, Bela and Nayak, Chetan},
  journal = {Phys. Rev. Lett.},
  volume = {117},
  issue = {9},
  pages = {090402},
  numpages = {5},
  year = {2016},
  month = {Aug},
  publisher = {American Physical Society},
  doi = {10.1103/PhysRevLett.117.090402},
  url = {https://link.aps.org/doi/10.1103/PhysRevLett.117.090402}
}

@article{von2016,
  title = {Phase structure of one-dimensional interacting Floquet systems. II. Symmetry-broken phases},
  author = {von Keyserlingk, C. W. and Sondhi, S. L.},
  journal = {Phys. Rev. B},
  volume = {93},
  issue = {24},
  pages = {245146},
  numpages = {11},
  year = {2016},
  month = {Jun},
  publisher = {American Physical Society},
  doi = {10.1103/PhysRevB.93.245146},
  url = {https://link.aps.org/doi/10.1103/PhysRevB.93.245146}
}

@article{Ponte2015,
  title = {Periodically driven ergodic and many-body localized quantum systems},
  volume = {353},
  ISSN = {0003-4916},
  url = {http://dx.doi.org/10.1016/j.aop.2014.11.008},
  DOI = {10.1016/j.aop.2014.11.008},
  journal = {Annals of Physics},
  publisher = {Elsevier BV},
  author = {Ponte,  Pedro and Chandran,  Anushya and Papić,  Z. and Abanin,  Dmitry A.},
  year = {2015},
  month = feb,
  pages = {196–204}
}

@article{von2016b,
  title = {Absolute stability and spatiotemporal long-range order in Floquet systems},
  author = {von Keyserlingk, C. W. and Khemani, Vedika and Sondhi, S. L.},
  journal = {Phys. Rev. B},
  volume = {94},
  issue = {8},
  pages = {085112},
  numpages = {11},
  year = {2016},
  month = {Aug},
  publisher = {American Physical Society},
  doi = {10.1103/PhysRevB.94.085112},
  url = {https://link.aps.org/doi/10.1103/PhysRevB.94.085112}
}

@article{Serbyn2013,
  title = {Local Conservation Laws and the Structure of the Many-Body Localized States},
  author = {Serbyn, Maksym and Papi\ifmmode \acute{c}\else \'{c}\fi{}, Z. and Abanin, Dmitry A.},
  journal = {Phys. Rev. Lett.},
  volume = {111},
  issue = {12},
  pages = {127201},
  numpages = {5},
  year = {2013},
  month = {Sep},
  publisher = {American Physical Society},
  doi = {10.1103/PhysRevLett.111.127201},
  url = {https://link.aps.org/doi/10.1103/PhysRevLett.111.127201}
}

@article{Huse2014,
  title = {Phenomenology of fully many-body-localized systems},
  author = {Huse, David A. and Nandkishore, Rahul and Oganesyan, Vadim},
  journal = {Phys. Rev. B},
  volume = {90},
  issue = {17},
  pages = {174202},
  numpages = {5},
  year = {2014},
  month = {Nov},
  publisher = {American Physical Society},
  doi = {10.1103/PhysRevB.90.174202},
  url = {https://link.aps.org/doi/10.1103/PhysRevB.90.174202}
}

@article{Fleckenstein2021,
  title = {Thermalization and prethermalization in periodically kicked quantum spin chains},
  author = {Fleckenstein, Christoph and Bukov, Marin},
  journal = {Phys. Rev. B},
  volume = {103},
  issue = {14},
  pages = {144307},
  numpages = {26},
  year = {2021},
  month = {Apr},
  publisher = {American Physical Society},
  doi = {10.1103/PhysRevB.103.144307},
  url = {https://link.aps.org/doi/10.1103/PhysRevB.103.144307}
}

@article{Ho2023,
  title = {Quantum and classical Floquet prethermalization},
  volume = {454},
  ISSN = {0003-4916},
  url = {http://dx.doi.org/10.1016/j.aop.2023.169297},
  DOI = {10.1016/j.aop.2023.169297},
  journal = {Annals of Physics},
  publisher = {Elsevier BV},
  author = {Ho,  Wen Wei and Mori,  Takashi and Abanin,  Dmitry A. and Dalla Torre,  Emanuele G.},
  year = {2023},
  month = jul,
  pages = {169297}
}

@article{agarwal2025b,
  title = {Quantum sensing with ultracold simulators in lattice and ensemble systems: A review},
  ISSN = {1793-6586},
  url = {http://dx.doi.org/10.1142/S0129183125430065},
  DOI = {10.1142/s0129183125430065},
  journal = {International Journal of Modern Physics C},
  publisher = {World Scientific Pub Co Pte Ltd},
  author = {Agarwal,  Keshav Das and Mondal,  Sayan and Sahoo,  Ayan and Rakshit,  Debraj and Sen(De),  Aditi and Sen,  Ujjwal},
  year = {2025},
  month = dec 
}

@article{Mishra2019,
  title = {Quantum correlations in periodically driven spin chains: Revivals and steady-state properties},
  volume = {491},
  ISSN = {0304-8853},
  url = {http://dx.doi.org/10.1016/j.jmmm.2019.165546},
  DOI = {10.1016/j.jmmm.2019.165546},
  journal = {Journal of Magnetism and Magnetic Materials},
  publisher = {Elsevier BV},
  author = {Mishra,  Utkarsh and Prabhu,  R. and Rakshit,  Debraj},
  year = {2019},
  month = dec,
  pages = {165546}
}

@article{Ghosh2020b,
  title = {Enhancement in the performance of a quantum battery by ordered and disordered interactions},
  author = {Ghosh, Srijon and Chanda, Titas and Sen(De), Aditi},
  journal = {Phys. Rev. A},
  volume = {101},
  issue = {3},
  pages = {032115},
  numpages = {11},
  year = {2020},
  month = {Mar},
  publisher = {American Physical Society},
  doi = {10.1103/PhysRevA.101.032115},
  url = {https://link.aps.org/doi/10.1103/PhysRevA.101.032115}
}

@article{Ghosh2021c,
  title = {Fast charging of a quantum battery assisted by noise},
  author = {Ghosh, Srijon and Chanda, Titas and Mal, Shiladitya and Sen(De), Aditi},
  journal = {Phys. Rev. A},
  volume = {104},
  issue = {3},
  pages = {032207},
  numpages = {12},
  year = {2021},
  month = {Sep},
  publisher = {American Physical Society},
  doi = {10.1103/PhysRevA.104.032207},
  url = {https://link.aps.org/doi/10.1103/PhysRevA.104.032207}
}

@article{konar2022b,
  title = {Quantum battery with ultracold atoms: Bosons versus fermions},
  author = {Konar, Tanoy Kanti and Lakkaraju, Leela Ganesh Chandra and Ghosh, Srijon and Sen(De), Aditi},
  journal = {Phys. Rev. A},
  volume = {106},
  issue = {2},
  pages = {022618},
  numpages = {12},
  year = {2022},
  month = {Aug},
  publisher = {American Physical Society},
  doi = {10.1103/PhysRevA.106.022618},
  url = {https://link.aps.org/doi/10.1103/PhysRevA.106.022618}
}

@article{Bhatta2024,
  title = {Noncompletely Positive Quantum Maps Enable Efficient Local Energy Extraction in Batteries},
  author = {Bhattacharyya, Aparajita and Sen, Kornikar and Sen, Ujjwal},
  journal = {Phys. Rev. Lett.},
  volume = {132},
  issue = {24},
  pages = {240401},
  numpages = {7},
  year = {2024},
  month = {Jun},
  publisher = {American Physical Society},
  doi = {10.1103/PhysRevLett.132.240401},
  url = {https://link.aps.org/doi/10.1103/PhysRevLett.132.240401}
}

@article{chaki2023,
  title={Auxiliary-assisted stochastic energy extraction from quantum batteries},
  author={Chaki, Paranjoy and Bhattacharyya, Aparajita and Sen, Kornikar and Sen, Ujjwal},
  journal={arXiv:2307.16856},
  url = {https://arxiv.org/abs/2307.16856},
  year={2023}
}

@article{sen2023noisy,
  title={Noisy quantum batteries},
  author={Sen, Kornikar and Sen, Ujjwal},
  journal={arXiv:2302.07166},
  url = {https://arxiv.org/abs/2302.07166},
  year={2023}
}

@article{PhysRevB.111.024315,
  title = {Prethermal Floquet time crystals in chiral multiferroic chains and applications as quantum sensors of AC fields},
  author = {Shukla, Rohit Kumar and Chotorlishvili, Levan and Mishra, Sunil K. and Iemini, Fernando},
  journal = {Phys. Rev. B},
  volume = {111},
  issue = {2},
  pages = {024315},
  numpages = {10},
  year = {2025},
  month = {Jan},
  publisher = {American Physical Society},
  doi = {10.1103/PhysRevB.111.024315},
  url = {https://link.aps.org/doi/10.1103/PhysRevB.111.024315}
}

@article{shukla2025optimizing,
  title={Optimizing quantum battery performance by reducing battery influence in charging dynamics},
  author={Shukla, Rohit Kumar and Kumar, Rajiv and Sen, Ujjwal and Mishra, Sunil K},
  journal={arXiv:2505.08029},
  url = {https://arxiv.org/abs/2505.08029},
  year={2025}
}

@article{Autti2020,
  title = {AC Josephson effect between two superfluid time crystals},
  volume = {20},
  ISSN = {1476-4660},
  url = {http://dx.doi.org/10.1038/s41563-020-0780-y},
  DOI = {10.1038/s41563-020-0780-y},
  number = {2},
  journal = {Nature Materials},
  publisher = {Springer Science and Business Media LLC},
  author = {Autti,  S. and Heikkinen,  P. J. and M\"{a}kinen,  J. T. and Volovik,  G. E. and Zavjalov,  V. V. and Eltsov,  V. B.},
  year = {2020},
  month = aug,
  pages = {171–174}
}

@article{PhysRevX.12.031037,
  title = {Discrete Time-Crystalline Response Stabilized by Domain-Wall Confinement},
  author = {Collura, Mario and De Luca, Andrea and Rossini, Davide and Lerose, Alessio},
  journal = {Phys. Rev. X},
  volume = {12},
  issue = {3},
  pages = {031037},
  numpages = {34},
  year = {2022},
  month = {Sep},
  publisher = {American Physical Society},
  doi = {10.1103/PhysRevX.12.031037},
  url = {https://link.aps.org/doi/10.1103/PhysRevX.12.031037}
}

@article{PhysRevLett.120.215301,
  title = {Observation of a Time Quasicrystal and Its Transition to a Superfluid Time Crystal},
  author = {Autti, S. and Eltsov, V. B. and Volovik, G. E.},
  journal = {Phys. Rev. Lett.},
  volume = {120},
  issue = {21},
  pages = {215301},
  numpages = {5},
  year = {2018},
  month = {May},
  publisher = {American Physical Society},
  doi = {10.1103/PhysRevLett.120.215301},
  url = {https://link.aps.org/doi/10.1103/PhysRevLett.120.215301}
}

@article{sarkar2025fluctuation,
  title={Fluctuation in energy extraction from quantum batteries: How open should the system be to control it?},
  author={Sarkar, Anindita and Chaki, Paranjoy and Ghosh, Priya and Sen, Ujjwal},
  journal={arXiv:2505.16851},
  url = {https://arxiv.org/abs/2505.16851},
  year={2025}
}

@article{debnath2025localization,
  title={Localization from Infinitesimal Kinetic Grading: Critical Scaling and Kibble-Zurek Universality},
  author={Debnath, Argha and Sahoo, Ayan and Rakshit, Debraj},
  journal={arXiv:2512.15795},
  url = {https://arxiv.org/abs/2512.15795},
  year={2025}
}

@article{PhysRevLett.98.090401,
  title = {Generalized Limits for Single-Parameter Quantum Estimation},
  author = {Boixo, Sergio and Flammia, Steven T. and Caves, Carlton M. and Geremia, JM},
  journal = {Phys. Rev. Lett.},
  volume = {98},
  issue = {9},
  pages = {090401},
  numpages = {4},
  year = {2007},
  month = {Feb},
  publisher = {American Physical Society},
  doi = {10.1103/PhysRevLett.98.090401},
  url = {https://link.aps.org/doi/10.1103/PhysRevLett.98.090401}
}

@article{PhysRevLett.92.207901,
  title = {Effective Quantum Spin Systems with Trapped Ions},
  author = {Porras, D. and Cirac, J. I.},
  journal = {Phys. Rev. Lett.},
  volume = {92},
  issue = {20},
  pages = {207901},
  numpages = {4},
  year = {2004},
  month = {May},
  publisher = {American Physical Society},
  doi = {10.1103/PhysRevLett.92.207901},
  url = {https://link.aps.org/doi/10.1103/PhysRevLett.92.207901}
}

@article{Islam2013,
  title = {Emergence and Frustration of Magnetism with Variable-Range Interactions in a Quantum Simulator},
  volume = {340},
  ISSN = {1095-9203},
  url = {http://dx.doi.org/10.1126/science.1232296},
  DOI = {10.1126/science.1232296},
  number = {6132},
  journal = {Science},
  publisher = {American Association for the Advancement of Science (AAAS)},
  author = {Islam,  R. and Senko,  C. and Campbell,  W. C. and Korenblit,  S. and Smith,  J. and Lee,  A. and Edwards,  E. E. and Wang,  C.-C. J. and Freericks,  J. K. and Monroe,  C.},
  year = {2013},
  month = may,
  pages = {583–587}
}

@article{RevModPhys.93.025001,
  title = {Programmable quantum simulations of spin systems with trapped ions},
  author = {Monroe, C. and Campbell, W. C. and Duan, L.-M. and Gong, Z.-X. and Gorshkov, A. V. and Hess, P. W. and Islam, R. and Kim, K. and Linke, N. M. and Pagano, G. and Richerme, P. and Senko, C. and Yao, N. Y.},
  journal = {Rev. Mod. Phys.},
  volume = {93},
  issue = {2},
  pages = {025001},
  numpages = {57},
  year = {2021},
  month = {Apr},
  publisher = {American Physical Society},
  doi = {10.1103/RevModPhys.93.025001},
  url = {https://link.aps.org/doi/10.1103/RevModPhys.93.025001}
}

@article{Frey2022,
  title = {Realization of a discrete time crystal on 57 qubits of a quantum computer},
  volume = {8},
  ISSN = {2375-2548},
  url = {http://dx.doi.org/10.1126/sciadv.abm7652},
  DOI = {10.1126/sciadv.abm7652},
  number = {9},
  journal = {Science Advances},
  publisher = {American Association for the Advancement of Science (AAAS)},
  author = {Frey,  Philipp and Rachel,  Stephan},
  year = {2022},
  month = Mar 
}

@article{Paulino2025,
  title = {Thermodynamics of coupled time crystals with an application to energy storage},
  volume = {11},
  ISSN = {2058-9565},
  url = {http://dx.doi.org/10.1088/2058-9565/ae186c},
  DOI = {10.1088/2058-9565/ae186c},
  number = {1},
  journal = {Quantum Science and Technology},
  publisher = {IOP Publishing},
  author = {Paulino,  Paulo J and Cabot,  Albert and De Chiara,  Gabriele and Antezza,  Mauro and Lesanovsky,  Igor and Carollo,  Federico},
  year = {2025},
  month = Nov,
  pages = {015003}
}

@article{Mukherjee2024,
  title = {The promises and challenges of many-body quantum technologies: A focus on quantum engines},
  volume = {15},
  ISSN = {2041-1723},
  url = {http://dx.doi.org/10.1038/s41467-024-47638-1},
  DOI = {10.1038/s41467-024-47638-1},
  number = {1},
  journal = {Nature Communications},
  publisher = {Springer Science and Business Media LLC},
  author = {Mukherjee,  Victor and Divakaran,  Uma},
  year = {2024},
  month = Apr 
}

@article{PhysRevLett.124.170602,
  title = {Nonequilibrium Quantum Many-Body Rydberg Atom Engine},
  author = {Carollo, Federico and Gambetta, Filippo M. and Brandner, Kay and Garrahan, Juan P. and Lesanovsky, Igor},
  journal = {Phys. Rev. Lett.},
  volume = {124},
  issue = {17},
  pages = {170602},
  numpages = {7},
  year = {2020},
  month = {Apr},
  publisher = {American Physical Society},
  doi = {10.1103/PhysRevLett.124.170602},
  url = {https://link.aps.org/doi/10.1103/PhysRevLett.124.170602}
}

@article{Ghosh26,
  title         = {Journey in Quantum Metrology and Sensing from Foundations to Applications: A Review},
  author        = {Ghosh, Priya and Konar, Tanoy Kanti and Rakshit, Debraj and Sen(De), Aditi and Sen, Ujjwal},
  journal       = {arXiv preprint arXiv:2605.21702},
  year          = {2026},
  eprint        = {2605.21702},
  archivePrefix = {arXiv},
  primaryClass  = {quant-ph}
}

@article{DAlessio2014LongTime,
  title   = {Long-Time Behavior of Isolated Periodically Driven Interacting Lattice Systems},
  author  = {D'Alessio, Luca and Rigol, Marcos},
  journal = {Phys. Rev. X},
  volume  = {4},
  pages   = {041048},
  year    = {2014},
  doi     = {10.1103/PhysRevX.4.041048}
}

@article{Giraud2022Probing,
  title   = {Probing Symmetries of Quantum Many-Body Systems through Gap Ratio Statistics},
  author  = {Giraud, Olivier and Mac{\'e}, Nicolas and Vernier, {\'E}ric and Alet, Fabien},
  journal = {Phys. Rev. X},
  volume  = {12},
  pages   = {011006},
  year    = {2022},
  doi     = {10.1103/PhysRevX.12.011006}
}

@article{Atas2013Distribution,
  title   = {Distribution of the Ratio of Consecutive Level Spacings in Random Matrix Ensembles},
  author  = {Atas, Y. Y. and Bogomolny, E. and Giraud, O. and Roux, G.},
  journal = {Phys. Rev. Lett.},
  volume  = {110},
  pages   = {084101},
  year    = {2013},
  doi     = {10.1103/PhysRevLett.110.084101}
}

\end{document}